\documentclass[%
 reprint,
superscriptaddress,
amsmath,amssymb,
aps,
prl,
]{revtex4-1}

\usepackage[colorlinks=true, allcolors=blue]{hyperref}
\usepackage{color}
\usepackage{natbib}
\usepackage{graphicx}
\usepackage[utf8]{inputenc}
\usepackage{accents}
\usepackage[super]{nth}
\usepackage{upgreek}
\usepackage[normalem]{ulem}

\begin{document}

\title{Precision measurement of the Lamb shift in Muonium}


\author{B.~Ohayon}
\author{G.~Janka}
\author{I.~Cortinovis}
\author{Z.~Burkley}
\author{L.~de Sousa Borges}
\author{E.~Depero}
\affiliation{
Institute for Particle Physics and Astrophysics, ETH Z\"urich, CH-8093 Z\"urich, Switzerland
}
\author{A.~Golovizin}
\affiliation{P.N. Lebedev Physical Institute, 53 Leninsky prospekt., Moscow 119991, Russia
}
\author{X.~Ni}
\author{Z.~Salman}
\author{A.~Suter}
\affiliation{Laboratory for Muon Spin Spectroscopy, Paul Scherrer Institute, CH-5232 Villigen PSI, Switzerland}
\author{C.~Vigo}
\affiliation{
Institute for Particle Physics and Astrophysics, ETH Z\"urich, CH-8093 Z\"urich, Switzerland
}
\author{T.~Prokscha}
\affiliation{Laboratory for Muon Spin Spectroscopy, Paul Scherrer Institute, CH-5232 Villigen PSI, Switzerland}
\author{P.~Crivelli}\email{Corresponding author: crivelli@phys.ethz.ch}
\affiliation{
Institute for Particle Physics and Astrophysics, ETH Z\"urich, CH-8093 Z\"urich, Switzerland
} 
\collaboration{Mu-MASS collaboration}

\begin{abstract}

We report a new measurement of the $n=2$ Lamb shift in Muonium using microwave spectroscopy.
Our result of $1047.2(2.3)_\textrm{stat}(1.1)_\textrm{syst}$\,MHz comprises an order of magnitude improvement upon the previous best measurement.
This value matches the theoretical calculation within one standard deviation allowing us to set limits on CPT violation in the muonic sector, as well as on new physics coupled to muons and electrons which could provide an explanation of the muon $g-2$ anomaly.

\end{abstract}

\maketitle

The classical Lamb shift (LS) is the difference between the $2S_{1/2}$ and $2P_{1/2}$ levels of a hydrogenic atom. 
It stems in the most part from the so called electron self-energy, i.e the possibility for the orbiting particle  to emit and re-absorb a virtual photon \cite{1939-Weiss}. This effect is not accounted for by the Dirac theory.
Following the clear identification of the LS in hydrogen by Lamb and Retherford \cite{1947-Lamb}, Bethe completed the first successful calculation of the interaction of an electron with the radiation field \cite{1947-Bethe}, finding agreement with the experimental result.
The discrepancy between the then-prevailing theory and measurements of the LS, together with that of the electron anomalous moment  \cite{1948-Schwin}, had found unambiguous evidence for what was at the time new physics. These findings lead to the development of modern Quantum Electrodynamics (QED).

In hydrogen, the increasing precision in LS measurements, as well as more advanced theoretical calculations, have made their comparison susceptible to the proton charge distribution.
Various discrepancies in the proton charge radius determination have appeared already in the 1980's \cite{1988-Disc}. They peaked when another LS measurement, that of muonic hydrogen \cite{2010-Pohl}, returned a significantly smaller radius than the accepted one. This is the so called proton radius puzzle \cite{2013-Puzzle}, with recent measurements arguably nearing its resolution \cite{2020-Puzzle,2020-PuzzleK,2021-Puzzle}.
This saga embodies the fact that when dealing with measurements involving hadronic particles such as protons, at a certain precision we learn more about their internal structure than about the consistency of QED.

In leptonic systems such as Muonium (M), the bound state of a positive muon and an electron; and Positronium, made of an electron and a positron; hadronic effects are entering only as loop corrections, and there are no finite-size contributions. Precision spectroscopy of such exotic atoms offers a clean arrow pointing at bound-state QED tests and an excellent probe for numerous scenarios beyond the Standard Model. These include Lorentz/CPT violations \cite{2014-Vargas}, new muonic forces \cite{Karshenboim:2014tka}, Dark Sectors \cite{2019-Dark}, the effect of gravity on anti-matter via the gravitational Redshift \cite{Karshenboim:2008zj} and to search for highly singular neutrino-like forces \cite{Stadnik:2017yge}.
However, despite their simplicity, these systems exhibit unique experimental challenges.

Positronium has a very short lifetime due to annihilation, resulting in broad resonances and limiting the available statistics.
Being the lightest known atom, Positronium travels at high velocities, even at room temperature, creating a tremendous experimental challenge.
Nevertheless, the Positronium $n=2$ fine structure was measured recently by microwave spectroscopy with sub MHz precision \cite{2020-PS}. Interestingly, this result differs by $4.5\,\sigma$ from the theoretical prediction and warrants further studies \cite{2021-Ps}.

With a longer lifetime of $2.2\,\upmu$s limited by the muon decay, and a larger mass, M constitutes a promising system for spectroscopic measurements \cite{2016-Jung,2021-MSPEC}.
The main challenge for obtaining M in vacuum, and especially in the long-lived 2S state, is that surface muons have a high kinetic energy of $4.1\,$MeV.

Previous experiments measuring the LS in M relied on a degrader to reduce the beam energy, creating a highly diffuse M distribution \cite{1981-M}. This resulted in low statistics and caused a large muon-related background. Nevertheless, the LS in M was measured to be $1070^{+12}_{-15}$ at TRIUMF \cite{1984-TRIUMF}, and $1042^{+21}_{-23}$ at LAMPF \cite{1990-LAMPF}. The main limitation for both measurements was the lack of a high quality, low energy, positive muon beam.
Such a beam is available today at the Low Energy Muon (LEM) beamline at the Paul Scherrer Institute (PSI). LEM is a unique beamline, used primarily for muon spin rotation ($\mu$SR) experiments \cite{2000-MuSR}, which provides a pure $\mu^+$ beam with selectable energy between $1$ and $30\,$keV \cite{2008-uE4}. 

Following our demonstration of the production of an intense M$_{2S}$ beam at the LEM \cite{2020-2S}, we report here on a precision measurement of the Lamb shift in Muonium comprising an order of magnitude  improvement over the state of the art. 

 \begin{figure*}[!ht]
    \centering
    \includegraphics[width=1.00\textwidth]{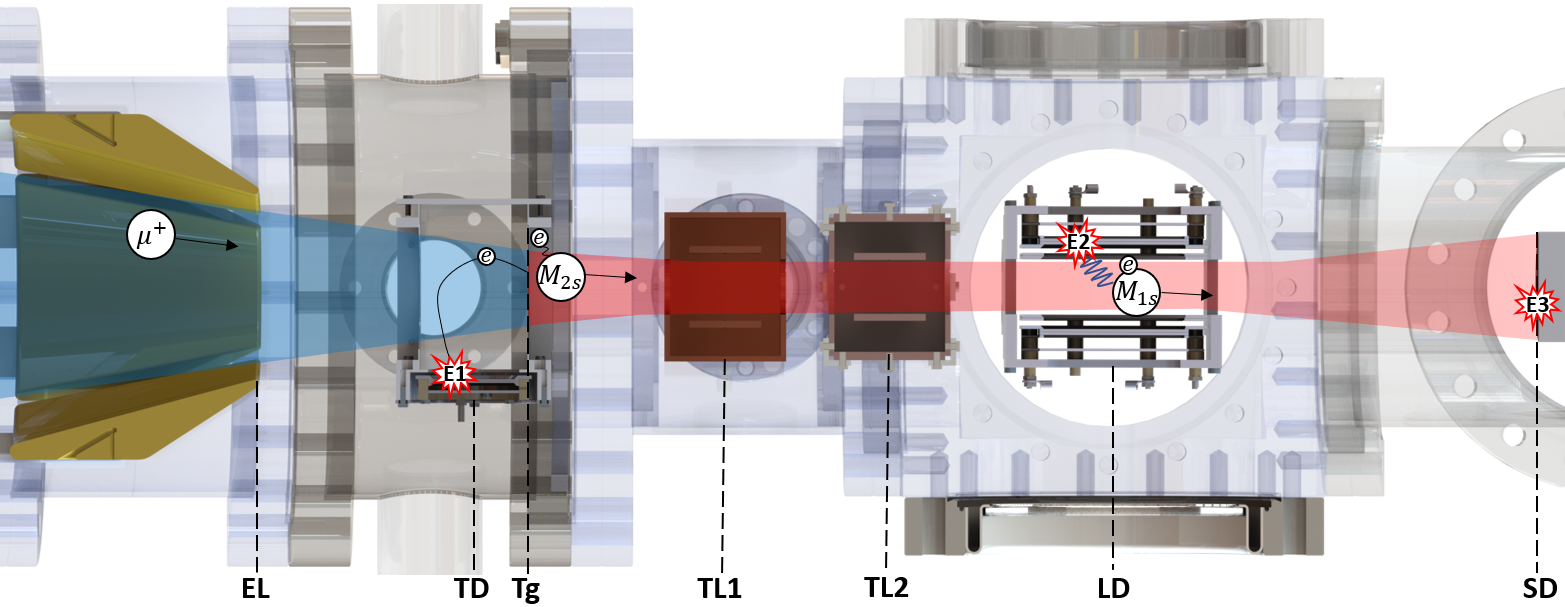}
    \caption{
Main elements of the experimental system. 
EL - conical electrostatic lens, TD - Tagging detector, Tg - Carbon Foil Target, TL - Transmission line, LD - Lyman-alpha detector, SD - Stop detector.
The normalization signal is given by the coincidence between an event in TD (E1) and SD (E3) within the expected time of flight, while a valid event includes also a reading in LD (E2).
    }
    \label{fig:exp_schematics}
\end{figure*}

In our experiment, surface muons ($\sim 2\times 10^8$ $\mu^+$/s, momentum $28\,$MeV/c) from the  $\upmu$E4 beamline \cite{2008-uE4} are implanted into a neon moderator from which they are re-emitted as epithermal muons with an energy of roughly $20\,$eV. These are accelerated by a DC voltage of $7.5\,$kV and transported electrostatically. 
An $E \times B$ filter selects only muons while rejecting other particles such as protons and neon atoms from the moderator \cite{2012-SpinRot}.
The beam is guided to our apparatus depicted in Fig. \ref{fig:exp_schematics}, where a segmented conical electrostatic lens \cite{xiao:2017nst} focuses it onto a $10\,$nm-thick grounded carbon foil target. The foil serves a double purpose: producing M, and tagging the incoming muons.

When passing the foil, a particle has a high probability of back-scattering an electron \cite{2016-Foils}. This electron is guided by an electric field and collected on a microchannel plate (MCP) detector, thus tagging the incoming muons on an event-by-event basis.
We measured the tagging efficiency, through coincidence measurement with another tagging detector, to be $67(3)\%$. The measured rate at the tagging detector was $13-15\,$kHz, out of which we estimate that $9.3(4)\,$kHz is the tagged muon rate at the foil, and the rest is beam-related background.
%
For an incident energy of $7.5\,$keV, $43(2)\%$ of the muons that traversed the foil produce Muonium, out of which $11(4)\%$ are in long-lived excited states \cite{2020-2S}.
In the absence of electromagnetic fields, the $2S$ M radiative lifetime is $0.12\,$s, which far exceeds the muon lifetime and any other time scale in this experiment.

Within the detection region, the beam encounters a static electric field of $250\,$V/cm, which mixes the $2S$ with the short-lived $2P$ levels and relaxes them within few nanoseconds to the ground state, emitting a Lyman-alpha photon at $122$ nm. This process is referred to as \textit{quenching}. These photons are detected by two coated MCPs placed around the quenching region. By simulating the field within this region, combined with the M position and energy distribution, we estimate that the total quenching and collection efficiency is around $40\%$.

\begin{figure}[!tb]
\centering
\includegraphics[width=1\columnwidth,trim={0 0 0 0},clip]{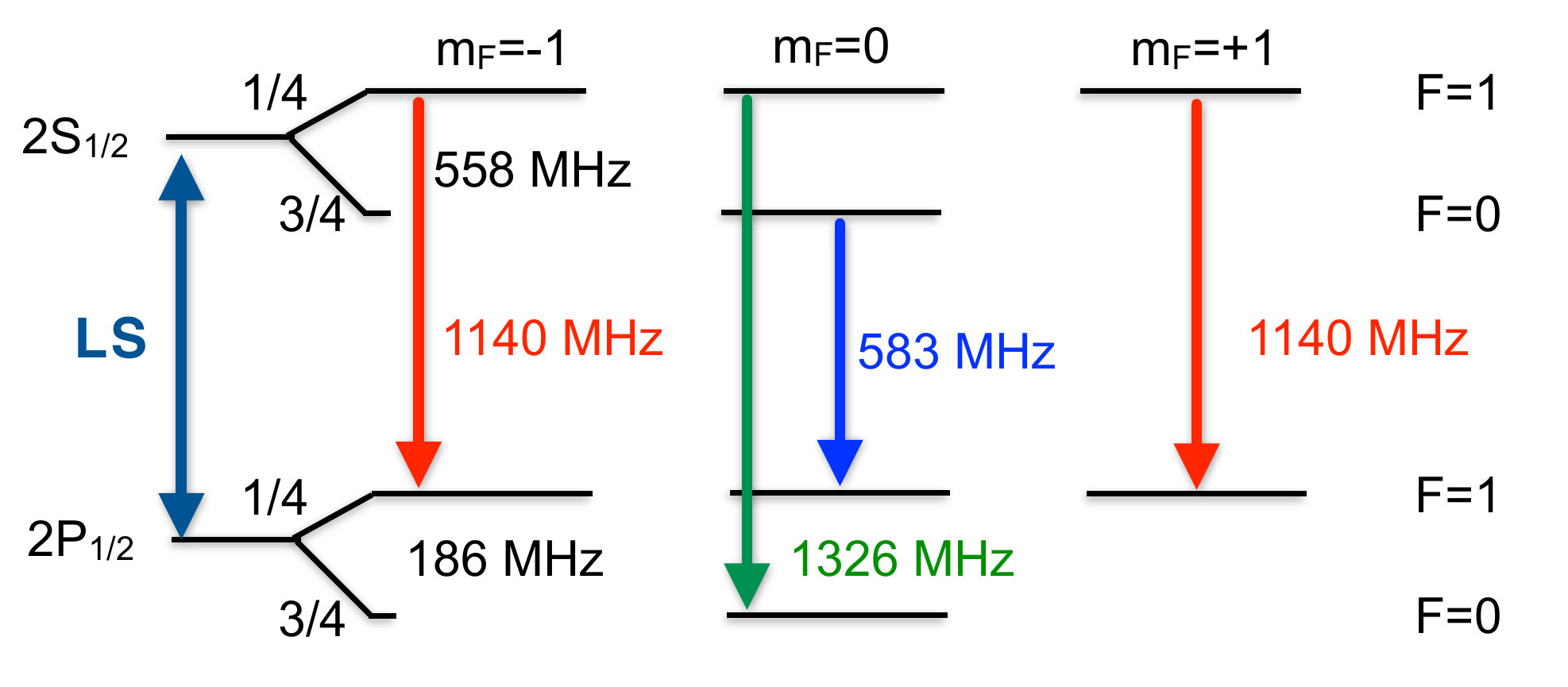}
\caption{\label{fig:LS_scheme}
Scheme of the M $2S_{1/2}$$-$$2P_{1/2}$ energy levels.
}
\end{figure}

Microwave (MW) radiation, resonant with one of the $^2S_{1/2}$$-$$^2P_{1/2}$ transitions (see Fig. \ref{fig:LS_scheme}), reduces the Lyman-alpha signal by depopulating M$_{2S}$ atoms before they reach the detection region. By scanning the MW around the different transitions, we measure the resonance shape from which we determine the line center and finally deduce the LS defined as the difference between the centroid $^2S_{1/2}$$-$$^2P_{1/2}$ binding energies averaged over all hyperfine levels  (as shown in Fig. \ref{fig:LS_scheme} or see also Eq.(1) of \cite{2019-Pachuski}).

For this purpose, we placed two balanced dual transmission lines (TL's) in the beam path. The design is similar to \cite{1986-LP}, in which a pair of parallel electrodes is driven $180$ degrees out of phase. In this configuration, M traveling in the center of the TL flies through a virtual ground plane which ensures that the field is linearly polarized perpendicular to the atomic beam.

At the end of the setup a position sensitive MCP detector acts as a beam monitor. This MCP is also used to measure the muon time-of-flight (TOF) and to trigger the data acquisition when a signal (events E3 in Fig. \ref{fig:exp_schematics}) is detected in coincidence with the tagging detector within $10\,\upmu$s.
We select TOFs that are compatible with those expected for muons at the given beam energy, taking into account losses in the foil. The coincidence rate (events E1 and E3 in Fig. \ref{fig:exp_schematics}) was $R_D=105\,$Hz and reduced slowly to $95\,$Hz within four days of continuous running due to aging of the neon moderator. $R_D$ is the sum of the muons, M, and a negligible background rate and is used as a normalization.
Our signal as a function of frequency is defined as $S(f)=R_T(f)/R_D(f)$, with $R_T$ the rate of triple-coincidences (events E1, E2 and E3 in Fig. \ref{fig:exp_schematics}). This signal corresponds to the probability of creating and detecting photons from quenched M in long-lived excited states per tagged muon that has reached SD.

To increase the signal to background and to reduce the lineshape uncertainty, we drive $86(5)\%$ of the $2S_{F=0}$ population to the ground state by applying a frequency of $580\,$MHz to TL2, while scanning the two $2S_{F=1}$ resonances in the range $900\mbox{--}1400\,$MHz with TL1.
Data taking took place for $48$ hours continuously, and constituted of cycles in which $9$ frequency points were measured (each point for 20 minutes), as well as one with TL1 off.

\begin{figure}[!t]
\centering
\includegraphics[width=1\columnwidth,trim={0 0 0 0},clip]{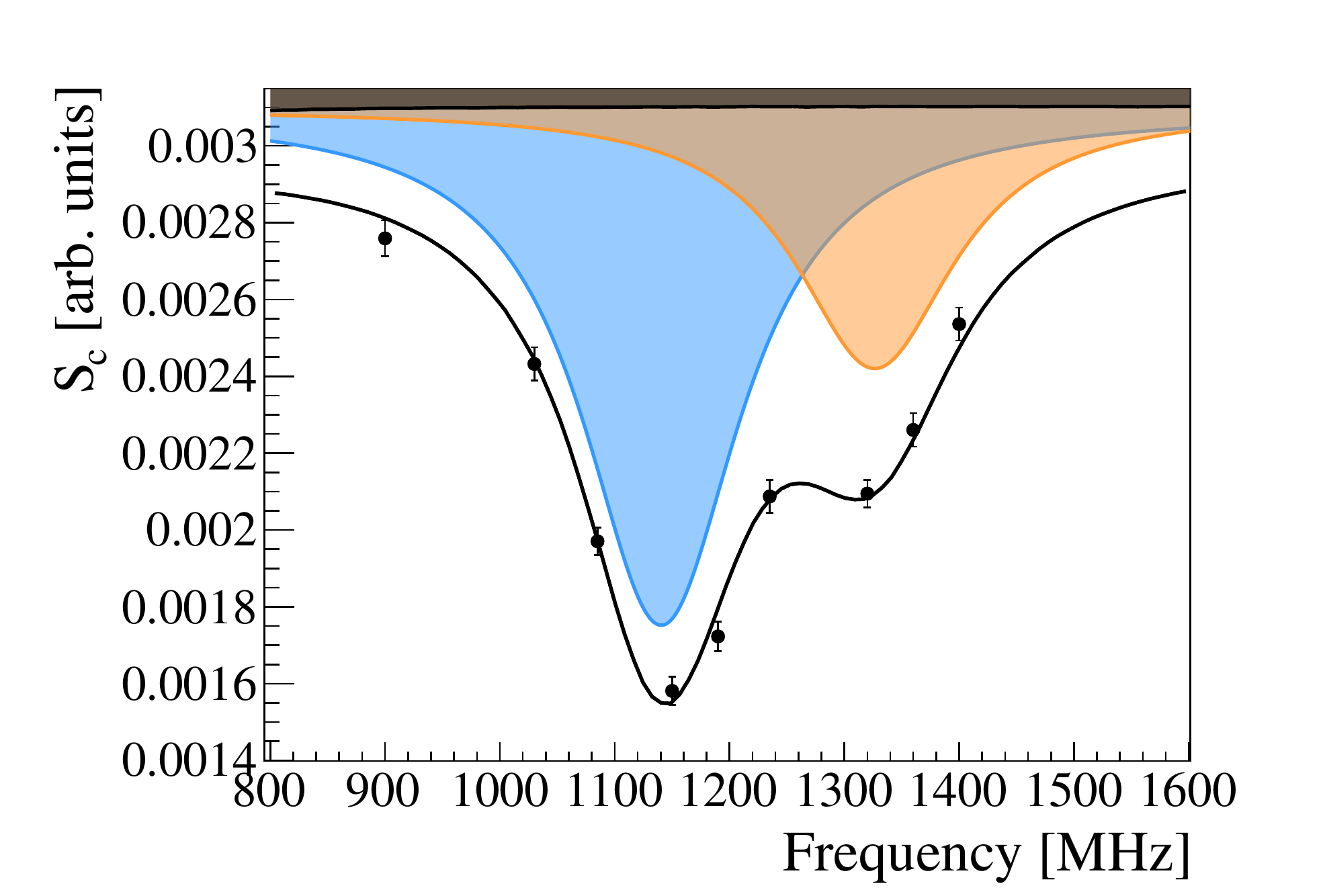}
\caption{\label{fig:LS_lineshape_combo_4S}
Measured resonance with the best line shape fit to the data (solid line). The MW off data point (not shown in the figure) lies at $(2.96\pm0.05)\times 10^{-3}$. The filled areas correspond to the individual contributions as described in the main text. 
}
\end{figure}

The total MW power was measured continuously outside the vacuum system. To determine the average power seen by the atoms in the TL center we measured the frequency-dependent power-loss and calibrated accordingly. We then applied a correction to the data-points in order to mimic data taken with a constant power of $29\,$W in the TL center.
The uncertainty in these corrections is added to the statistical one in quadrature.
The corrected signal is denoted $S_c(f)$ and is plotted in Fig. \ref{fig:LS_lineshape_combo_4S}. Its main features are two overlapping resonances corresponding to the two allowed transitions from the $2S_{F=1}$ hyperfine level.

To determine the line center from $S_c(f)$ we fit the data with line shapes obtained from a detailed Monte-Carlo simulation based on Geant 4 modeling of the LEM beamline \cite{sedlak:musrsim2012}, including the interaction with the thin carbon foil validated with experimental data \cite{Khaw:2015eya}. The same selection criteria for the M time-of-flight applied to the experimental data are used for the simulated events. To calculate the transition probability we use the optical Bloch equation for a 2-level system \cite{2006-Haas}.
The density matrix equations are integrated numerically for each atom's trajectory within the simulated time-dependent fields of the TL's. Since we do not employ the rotating wave approximation, the Bloch-Siegert shift is included in our line shape model. The MW phase is chosen randomly for each atom before entering the TL's so that it is averaged out.

The transitions of interest for our measurement are the
$2S_{F=0}$$-$$2P_{1/2,F=1}$ at $583\,$MHz, $2S_{F=1}$$-$$2P_{1/2,F=1}$ at $1140\,$MHz and the $2S_{F=1}$$-$$2P_{1/2,F=0}$ at $1326\,$MHz
(see Fig. \ref{fig:LS_scheme}).
The fitting procedure is done by $\chi^2$ minimization using the Minuit library within the ROOT package \cite{1997-ROOT}.
The free parameters of the fit are: a horizontal offset 
from the simulated $2S_{F=1}$$-$$2P_{1/2,F=1}$ resonance position, an overall magnitude, and a frequency-independent background parameter.
The resonance position obtained by the fitting is $1139.9(2.3)\,$MHz with a reduced chi square of $1.16$ for $7$ degrees of freedom. 

The compact nature of our setup, coupled with the high velocity of the beam, which travels on average $70$ ns between the foil and our detection system, makes us susceptible to contamination from other long-lived excited states which are populated by the beam-foil interaction and have transitions within our frequency range. The relevant transitions are between $4S-4P_{3/2}$ around $1240\,$MHz \cite{1976-Newton}.
From the horizontal offset of $5.9(6) \times10^{-4}$ obtained from the fit, we estimate the maximal contribution of all excited states to be $17\%$. Using the scaling factor of $1/n^3$ as in \cite{1979-Newton}, we get a maximal contribution of $5\%$ for 4S states. 
This is validated by keeping the magnitude of the 4S contribution as a free parameter in the fit, which returns a value compatible with zero within uncertainty. A dedicated measurement of the LS in hydrogen using protons available in the same beamline was performed, which supports our limit.
Therefore, to estimate this contribution, we fit the data adding 5\% 4S fraction and conservatively use the full  $1.0\,$MHz offset as the uncertainty to the line center.

To estimate the influence of the simulation input parameters on the resonance position, we scan them around their uncertainty and repeat the analysis. The ones that were found to give the largest contributions are:
\begin{enumerate}
\item  The MW field intensity which has an uncertainty of $6\%$ originating from the absolute calibration of the power meter ($3\%$) and from the mechanical tolerances in the TL construction ($5\%$). The estimated systematic uncertainty is at a level of $0.04\,$MHz.
It arises from the uncertainty in the AC Stark shift.

\item The M velocity distribution determined by the muon scattering in the thin C-foil.
As demonstrated with previous measurements, our simulation reproduces the TOF distribution very accurately with an uncertainty on the mean energy loss of a few $\%$ \cite{Khaw:2015eya}. This corresponds to a mean M energy of $E_M=5.7(2)\,$keV resulting in an uncertainty of $0.01\,$MHz. 

\item A misalignment of the TL axis with respect to the beam could lead to a residual first order Doppler shift. Taking into account the mechanical tolerances in the TL construction, we set an upper limit on the misalignment angle to be $30\,$mrad which would amount to a $0.32\,$MHz shift. 
\end{enumerate}
The above contributions are summarized in table \ref{tab:results}.

Other effects not included in the simulation can shift the central value.
The main one is an overestimation of the AC Stark shift as we did not include the $2P_{3/2}$ states in the Bloch equations. Nevertheless, this effect for the LS is well understood theoretically \cite{1976-BS} and validated experimentally \cite{1975-BS-EXP}, so that the AC Stark shift can be corrected by multiplying it with a factor $0.64$ resulting in $+0.26(2)\,$MHz as given in table \ref{tab:results}.
We evaluate the magnitude of several smaller systematic effects, namely the \nth{2}-order Doppler, motional Stark-shift from the Earth's magnetic field, and quantum interference-shift from the presence of $M_{3S}$ \cite{2017-QI}. These are given in table \ref{tab:results}.

Adding the various corrections, the determined frequency of the  $2S_{F=1}$$-$$2P_{1/2,F=1}$  transition is $1140.2(2.3)_\textrm{stat}(1.1)_\textrm{syst}$\,MHz and the corresponding LS is $1047.2(2.5)$\,MHz, where we added the statistical and systematic uncertainties in quadrature.
Our result is within one standard deviation from the theoretical value quoted in the literature of $1047.5(3)$\,MHz \cite{1990-LAMPF} (to be updated with recent bound state QED developments in hydrogen \cite{2019-Pachuski}) and a recent calculation based on effective field theory giving $1047.284(2)$\,MHz \cite{Peset:2015zga,2019-Dark}.

Since our result is in agreement with the theoretical calculations, we can use it to place stringent limits on new physics scenarios. Here we focus on possible Lorentz and CPT violation effects, and new bosons interacting with muons and electrons.
The M Lamb shift is sensitive to two of the isotropic nonrelativistic effective coefficients for Lorentz and CPT violation \cite{2014-Vargas}: namely $\accentset{\circ}{a}_{4}^{\mathrm{NR}}$ and $\accentset{\circ}{c}_{4}^{\mathrm{NR}}$. Taking conservatively $2\sigma$, we can set a bound on the linear combination:
\begin{equation}
    \left|\accentset{\circ}{a}_{4}^{\mathrm{NR}} + \accentset{\circ}{c}_{4}^{\mathrm{NR}} \right| < 1.7 \times 10^5 \text{ GeV\textsuperscript{-3}  },
\end{equation}
which translates into Table \ref{tab:newphysicsCPT}, when considering only one coefficient at a time to be non-zero. These bounds are of the same order as the current ones obtained from the measurement of the 1S$-$2S transition in M \cite{2000-1S2S}, and improve by an order of magnitude the previous bounds from the M Lamb Shift.

\begin{table}[t]

    \begin{ruledtabular} 
\begin{tabular}{lcc}
                & Central Value & Uncertainty\\
    Fitting              &  $1139.9$      &~~  $2.3$  \\
    4S contribution       &              & $<1.0$  \\
    MW-Beam alignment    &        & $<0.32$ \\
    MW field intensity      &        & $<0.04$  \\
    M velocity distribution          &        & $<0.01$  \\
    AC Stark $2P_{3/2}$      &~ $+0.26$ & $<0.02$ \\ 
    

    \nth{2}-order Doppler  &~ $+0.06$ & $<0.01$ \\ 
    Earth's Field          &        & $<0.05$ \\ 
    Quantum Interference   &        & $<0.04$ \\ 
 \hline
    $2S_{F=1}$$-$$2P_{1/2,F=1}$  & $1140.2$  & $2.5$ \\ 
        Hyperfine  & ~$-93.0$  & $0.0$ \\ 
 \hline
        Lamb Shift & $1047.2$  & $2.5$  \\ 
       Theoretical value \cite{2019-Dark} &~~ $1047.284$  &~ ~$0.002$ \\ 
    \end{tabular}
    \end{ruledtabular}
    \caption{Central values and uncertainty contributions in MHz.}
    \label{tab:results}
\end{table}

\begin{table}[!b]
{\renewcommand{\arraystretch}{2}%
\begin{tabular}{cccc}
\hline
\hline
Coefficient & & & Constraint \\ \hline
$\left|\accentset{\circ}{a}_{4}^{\mathrm{NR}}\right| $ &   &   & $< 1.7 \times 10^5$ GeV\textsuperscript{-3}       \\
$\left|\accentset{\circ}{c}_{4}^{\mathrm{NR}}\right| $ &   &   & $< 1.7 \times 10^5$ GeV\textsuperscript{-3}         \\ \hline
\end{tabular}
}
\caption{\label{tab:newphysicsCPT} Single constraints from the Lamb Shift measurement on isotropic nonrelativistic coefficients for CPT violation.}
\end{table}

\begin{figure}[!tb]
\centering
\includegraphics[width=1\columnwidth,trim={0 0 0 0},clip]{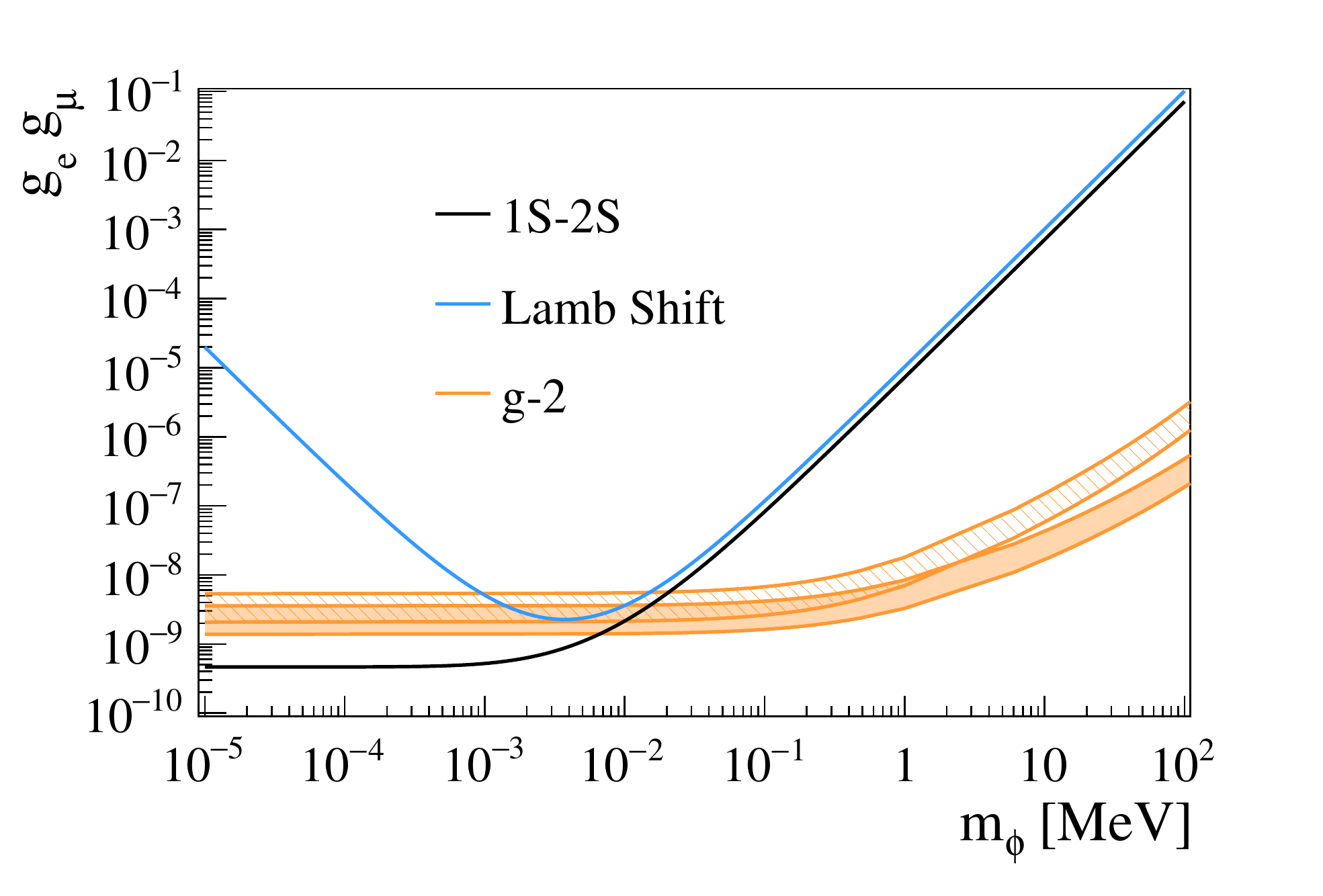}
    \caption{Constraints from M spectroscopy on $g^{s}_e, g^{s}_\mu$ as a function of the scalar/vector mass. The solid black line is the constraint from the M 1S$-$2S measurement \cite{2000-1S2S}  while the blue line is from the LS measurement presented here. The orange band represents the region suggested by the $g-2$ muon anomaly considering the lower bound from the measurement of the electron gyromagnetic factor for the scalar case, while the hatched region is for the vector one.}
\label{fig:scalar}
\end{figure}

A new scalar or a new vector gauge boson could provide an explanation of the muon $g-2$  anomaly \cite{ JEGERLEHNER20091, 2021-g-2,Balkin:2021rvh}. If this particle is carrying a dark force between electrons and muons, then M spectroscopy offers the possibility to search for it \cite{2019-Dark}.
For the scalar case, one has a Yukawa-like attractive potential of the form (see e.g. \cite{Fadeev:2018rfl}):

\begin{equation}
V_{ss}(\vec{r})=-g_e^s g^s_\mu \frac{e ^{-m_s r}}{4 \pi r},
\end{equation}
where $m_{s}$ is the scalar boson mass and $g^{s}_e,\,g^{s}_\mu$ are the coupling strengths to electrons and anti-muons, respectively. For small coupling strengths, the effect of such a potential can be calculated by applying perturbation theory. The vector potential can be found in \cite{Fadeev:2018rfl}.
In Fig. \ref{fig:scalar}, we present the sensitivity of Muonium spectroscopy to new physics. The constraints on $g^{s}_e,\,g^{s}_\mu$ as a function of the scalar/vector mass, which are nearly identical in the mass range considered here, are compared to the region favored by the $g-2$ muon anomaly \cite{Muong-2:2021ojo}, considering the bounds from the electron gyromagnetic factor \cite{Hanneke:2008tm}.  In fact, the experimental value of the electron anomalous magnetic moment is in agreement with the theoretical one when using as an input the recent new determinations of the fine structure constant \cite{Parker:2018vye,Morel:2020dww}.  We do not present results from experiments at the intensity frontier since those can be argued to be model dependent, i.e. to depend on their production or decay channels. An example is the recent results of the NA64 experiment placing stringent bounds on new bosons with the assumption that those would decay invisibly \cite{NA64:2021xzo}.
The combination of the LS and 1S$-$2S M measurements provides the most stringent laboratory constraint excluding that a new scalar/vector boson with a mass $<10\,$keV could contribute to the muon $g-2$ anomaly. 

To conclude, we reported a new measurement of the $n=2$ Lamb shift in Muonium using microwave spectroscopy. Our result of $1047.2(2.3)_\textrm{stat}(1.1)_\textrm{syst}$\,MHz comprises an order of magnitude of improvement upon the best previously determined value. As it agrees with the theoretical value within one standard deviation, we have set limits on CPT violation in the muonic sector, as well as new physics coupled to muons and electrons.

A major increase in statistics could be obtained if beam scattering by the foil would be considerably reduced. 
Recently, few layer graphene foils have been successfully used for both proton tagging and hydrogen production for space missions \cite{2014-GraphState}. These are $5\mbox{--}10$ times thinner than our currently used foil. As the scattering half angle is proportional to the foil thickness \cite{2020-Graph}, our statistics could be increased by a factor $15\mbox{--}25$. 
With an order of magnitude more events, the main systematic uncertainty of the experiment, originating from a possible contamination of the M$_{2S}$ beam of higher excited states, would become comparable with the statistical one. These states could be suppressed by introducing a weak quenching field between Tg and the TL's \cite{2019-Hessels}. Beam misalignment could be compensated by periodically reversing the MW direction, thus reducing the associated uncertainty by more than an order of magnitude \cite{1976-Newton}.
These straightforward improvements would increase the precision to $\sim 300\,$kHz level, which would significantly extend the reach of Muonium spectroscopy to search for new physics.

The realization of the  High intensity Muon Beam (HiMB) at PSI \cite{2017-HiMBMuCOOL} would open the way to push the accuracy of this experiment to its ultimate limit of few tens of kHz. This precision, along with other ongoing M spectroscopy experiments \cite{2018-MuMASS, 2020-MUSEUM,2021-MSPEC} would allow to fully probe the very interesting region of parameter space for a new scalar/vector boson with $m_\phi >300\,$keV suggested by the $g-2$ anomaly where astrophysical bounds do not apply \cite{2017-Stellar}.

\paragraph*{Acknowledgements} 
This work is based on experiments performed at the Swiss Muon Source S$\mu$S, Paul Scherrer Institute, Villigen, Switzerland.
This work is supported by the ERC consolidator grant 818053-Mu-MASS and the Swiss National Science Foundation under the grant 197346.
BO acknowledges support from the European Union’s Horizon 2020 research and innovation program under the Marie Skłodowska-Curie grant agreement No.~101019414, as well as ETH Zurich through a Career Seed Grant SEED-09 20-1.
We are grateful to D. Kirpichnikov, J. Northey, Y. Soreq and  A. Vargas for their help in setting the constraints from our measurement on new physics.  
\bibliography{references}

\begin{thebibliography}{57}%
\makeatletter
\providecommand \@ifxundefined [1]{%
 \@ifx{#1\undefined}
}%
\providecommand \@ifnum [1]{%
 \ifnum #1\expandafter \@firstoftwo
 \else \expandafter \@secondoftwo
 \fi
}%
\providecommand \@ifx [1]{%
 \ifx #1\expandafter \@firstoftwo
 \else \expandafter \@secondoftwo
 \fi
}%
\providecommand \natexlab [1]{#1}%
\providecommand \enquote  [1]{``#1''}%
\providecommand \bibnamefont  [1]{#1}%
\providecommand \bibfnamefont [1]{#1}%
\providecommand \citenamefont [1]{#1}%
\providecommand \href@noop [0]{\@secondoftwo}%
\providecommand \href [0]{\begingroup \@sanitize@url \@href}%
\providecommand \@href[1]{\@@startlink{#1}\@@href}%
\providecommand \@@href[1]{\endgroup#1\@@endlink}%
\providecommand \@sanitize@url [0]{\catcode `\\12\catcode `\$12\catcode
  `\&12\catcode `\#12\catcode `\^12\catcode `\_12\catcode `\%12\relax}%
\providecommand \@@startlink[1]{}%
\providecommand \@@endlink[0]{}%
\providecommand \url  [0]{\begingroup\@sanitize@url \@url }%
\providecommand \@url [1]{\endgroup\@href {#1}{\urlprefix }}%
\providecommand \urlprefix  [0]{URL }%
\providecommand \Eprint [0]{\href }%
\providecommand \doibase [0]{http://dx.doi.org/}%
\providecommand \selectlanguage [0]{\@gobble}%
\providecommand \bibinfo  [0]{\@secondoftwo}%
\providecommand \bibfield  [0]{\@secondoftwo}%
\providecommand \translation [1]{[#1]}%
\providecommand \BibitemOpen [0]{}%
\providecommand \bibitemStop [0]{}%
\providecommand \bibitemNoStop [0]{.\EOS\space}%
\providecommand \EOS [0]{\spacefactor3000\relax}%
\providecommand \BibitemShut  [1]{\csname bibitem#1\endcsname}%
\let\auto@bib@innerbib\@empty
\bibitem [{\citenamefont {Weisskopf}(1939)}]{1939-Weiss}%
  \BibitemOpen
  \bibfield  {author} {\bibinfo {author} {\bibfnamefont {V.~F.}\ \bibnamefont
  {Weisskopf}},\ }\href {\doibase 10.1103/PhysRev.56.72} {\bibfield  {journal}
  {\bibinfo  {journal} {Phys. Rev.}\ }\textbf {\bibinfo {volume} {56}},\
  \bibinfo {pages} {72} (\bibinfo {year} {1939})}\BibitemShut {NoStop}%
\bibitem [{\citenamefont {Lamb}\ and\ \citenamefont
  {Retherford}(1947)}]{1947-Lamb}%
  \BibitemOpen
  \bibfield  {author} {\bibinfo {author} {\bibfnamefont {W.~E.}\ \bibnamefont
  {Lamb}}\ and\ \bibinfo {author} {\bibfnamefont {R.~C.}\ \bibnamefont
  {Retherford}},\ }\href {\doibase 10.1103/PhysRev.72.241} {\bibfield
  {journal} {\bibinfo  {journal} {Phys. Rev.}\ }\textbf {\bibinfo {volume}
  {72}},\ \bibinfo {pages} {241} (\bibinfo {year} {1947})}\BibitemShut
  {NoStop}%
\bibitem [{\citenamefont {Bethe}(1947)}]{1947-Bethe}%
  \BibitemOpen
  \bibfield  {author} {\bibinfo {author} {\bibfnamefont {H.~A.}\ \bibnamefont
  {Bethe}},\ }\href {\doibase 10.1103/PhysRev.72.339} {\bibfield  {journal}
  {\bibinfo  {journal} {Phys. Rev.}\ }\textbf {\bibinfo {volume} {72}},\
  \bibinfo {pages} {339} (\bibinfo {year} {1947})}\BibitemShut {NoStop}%
\bibitem [{\citenamefont {Schwinger}(1948)}]{1948-Schwin}%
  \BibitemOpen
  \bibfield  {author} {\bibinfo {author} {\bibfnamefont {J.}~\bibnamefont
  {Schwinger}},\ }\href {\doibase 10.1103/PhysRev.73.416} {\bibfield  {journal}
  {\bibinfo  {journal} {Phys. Rev.}\ }\textbf {\bibinfo {volume} {73}},\
  \bibinfo {pages} {416} (\bibinfo {year} {1948})}\BibitemShut {NoStop}%
\bibitem [{\citenamefont {Grotch}(1988)}]{1988-Disc}%
  \BibitemOpen
  \bibfield  {author} {\bibinfo {author} {\bibfnamefont {H.}~\bibnamefont
  {Grotch}},\ }\href {\doibase 10.1088/0031-8949/1988/t21/016} {\bibfield
  {journal} {\bibinfo  {journal} {Physica Scripta}\ }\textbf {\bibinfo {volume}
  {T21}},\ \bibinfo {pages} {86} (\bibinfo {year} {1988})}\BibitemShut
  {NoStop}%
\bibitem [{\citenamefont {Pohl}\ \emph {et~al.}(2010)\citenamefont {Pohl},
  \citenamefont {Antognini}, \citenamefont {Nez}, \citenamefont {Amaro},
  \citenamefont {Biraben}, \citenamefont {Cardoso}, \citenamefont {Covita},
  \citenamefont {Dax}, \citenamefont {Dhawan}, \citenamefont {Fernandes} \emph
  {et~al.}}]{2010-Pohl}%
  \BibitemOpen
  \bibfield  {author} {\bibinfo {author} {\bibfnamefont {R.}~\bibnamefont
  {Pohl}}, \bibinfo {author} {\bibfnamefont {A.}~\bibnamefont {Antognini}},
  \bibinfo {author} {\bibfnamefont {F.}~\bibnamefont {Nez}}, \bibinfo {author}
  {\bibfnamefont {F.~D.}\ \bibnamefont {Amaro}}, \bibinfo {author}
  {\bibfnamefont {F.}~\bibnamefont {Biraben}}, \bibinfo {author} {\bibfnamefont
  {J.~M.}\ \bibnamefont {Cardoso}}, \bibinfo {author} {\bibfnamefont {D.~S.}\
  \bibnamefont {Covita}}, \bibinfo {author} {\bibfnamefont {A.}~\bibnamefont
  {Dax}}, \bibinfo {author} {\bibfnamefont {S.}~\bibnamefont {Dhawan}},
  \bibinfo {author} {\bibfnamefont {L.~M.}\ \bibnamefont {Fernandes}},  \emph
  {et~al.},\ }\href {\doibase https://doi.org/10.1038/nature09250} {\bibfield
  {journal} {\bibinfo  {journal} {Nature}\ }\textbf {\bibinfo {volume} {466}},\
  \bibinfo {pages} {213} (\bibinfo {year} {2010})}\BibitemShut {NoStop}%
\bibitem [{\citenamefont {Pohl}\ \emph {et~al.}(2013)\citenamefont {Pohl},
  \citenamefont {Gilman}, \citenamefont {Miller},\ and\ \citenamefont
  {Pachucki}}]{2013-Puzzle}%
  \BibitemOpen
  \bibfield  {author} {\bibinfo {author} {\bibfnamefont {R.}~\bibnamefont
  {Pohl}}, \bibinfo {author} {\bibfnamefont {R.}~\bibnamefont {Gilman}},
  \bibinfo {author} {\bibfnamefont {G.~A.}\ \bibnamefont {Miller}}, \ and\
  \bibinfo {author} {\bibfnamefont {K.}~\bibnamefont {Pachucki}},\ }\href
  {\doibase 10.1146/annurev-nucl-102212-170627} {\bibfield  {journal} {\bibinfo
   {journal} {Annual Review of Nuclear and Particle Science}\ }\textbf
  {\bibinfo {volume} {63}},\ \bibinfo {pages} {175} (\bibinfo {year}
  {2013})}\BibitemShut {NoStop}%
\bibitem [{\citenamefont {Ubachs}(2020)}]{2020-Puzzle}%
  \BibitemOpen
  \bibfield  {author} {\bibinfo {author} {\bibfnamefont {W.}~\bibnamefont
  {Ubachs}},\ }\href {\doibase 10.1126/science.abf0589} {\bibfield  {journal}
  {\bibinfo  {journal} {Science}\ }\textbf {\bibinfo {volume} {370}},\ \bibinfo
  {pages} {1033} (\bibinfo {year} {2020})}\BibitemShut {NoStop}%
\bibitem [{\citenamefont {Karr}\ \emph {et~al.}(2020)\citenamefont {Karr},
  \citenamefont {Marchand},\ and\ \citenamefont {Voutier}}]{2020-PuzzleK}%
  \BibitemOpen
  \bibfield  {author} {\bibinfo {author} {\bibfnamefont {J.-P.}\ \bibnamefont
  {Karr}}, \bibinfo {author} {\bibfnamefont {D.}~\bibnamefont {Marchand}}, \
  and\ \bibinfo {author} {\bibfnamefont {E.}~\bibnamefont {Voutier}},\ }\href
  {\doibase https://doi.org/10.1038/s42254-020-0229-x} {\bibfield  {journal}
  {\bibinfo  {journal} {Nature Reviews Physics}\ }\textbf {\bibinfo {volume}
  {2}},\ \bibinfo {pages} {601} (\bibinfo {year} {2020})}\BibitemShut {NoStop}%
\bibitem [{\citenamefont {Gao}\ and\ \citenamefont
  {Vanderhaeghen}(2021)}]{2021-Puzzle}%
  \BibitemOpen
  \bibfield  {author} {\bibinfo {author} {\bibfnamefont {H.}~\bibnamefont
  {Gao}}\ and\ \bibinfo {author} {\bibfnamefont {M.}~\bibnamefont
  {Vanderhaeghen}},\ }\href@noop {} {\enquote {\bibinfo {title} {The proton
  charge radius},}\ } (\bibinfo {year} {2021}),\ \Eprint
  {http://arxiv.org/abs/2105.00571} {arXiv:2105.00571 [hep-ph]} \BibitemShut
  {NoStop}%
\bibitem [{\citenamefont {Gomes}\ \emph {et~al.}(2014)\citenamefont {Gomes},
  \citenamefont {Kosteleck\'y},\ and\ \citenamefont {Vargas}}]{2014-Vargas}%
  \BibitemOpen
  \bibfield  {author} {\bibinfo {author} {\bibfnamefont {A.~H.}\ \bibnamefont
  {Gomes}}, \bibinfo {author} {\bibfnamefont {V.~A.}\ \bibnamefont
  {Kosteleck\'y}}, \ and\ \bibinfo {author} {\bibfnamefont {A.~J.}\
  \bibnamefont {Vargas}},\ }\href {\doibase 10.1103/PhysRevD.90.076009}
  {\bibfield  {journal} {\bibinfo  {journal} {Phys. Rev. D}\ }\textbf {\bibinfo
  {volume} {90}},\ \bibinfo {pages} {076009} (\bibinfo {year}
  {2014})}\BibitemShut {NoStop}%
\bibitem [{\citenamefont {Karshenboim}\ \emph {et~al.}(2014)\citenamefont
  {Karshenboim}, \citenamefont {McKeen},\ and\ \citenamefont
  {Pospelov}}]{Karshenboim:2014tka}%
  \BibitemOpen
  \bibfield  {author} {\bibinfo {author} {\bibfnamefont {S.~G.}\ \bibnamefont
  {Karshenboim}}, \bibinfo {author} {\bibfnamefont {D.}~\bibnamefont {McKeen}},
  \ and\ \bibinfo {author} {\bibfnamefont {M.}~\bibnamefont {Pospelov}},\
  }\href {\doibase 10.1103/PhysRevD.90.073004} {\bibfield  {journal} {\bibinfo
  {journal} {Phys. Rev. D}\ }\textbf {\bibinfo {volume} {90}},\ \bibinfo
  {pages} {073004} (\bibinfo {year} {2014})},\ \bibinfo {note} {[Addendum:
  Phys.Rev.D 90, 079905 (2014)]},\ \Eprint {http://arxiv.org/abs/1401.6154}
  {arXiv:1401.6154 [hep-ph]} \BibitemShut {NoStop}%
\bibitem [{\citenamefont {Frugiuele}\ \emph {et~al.}(2019)\citenamefont
  {Frugiuele}, \citenamefont {P\'erez-R\'{\i}os},\ and\ \citenamefont
  {Peset}}]{2019-Dark}%
  \BibitemOpen
  \bibfield  {author} {\bibinfo {author} {\bibfnamefont {C.}~\bibnamefont
  {Frugiuele}}, \bibinfo {author} {\bibfnamefont {J.}~\bibnamefont
  {P\'erez-R\'{\i}os}}, \ and\ \bibinfo {author} {\bibfnamefont
  {C.}~\bibnamefont {Peset}},\ }\href {\doibase 10.1103/PhysRevD.100.015010}
  {\bibfield  {journal} {\bibinfo  {journal} {Phys. Rev. D}\ }\textbf {\bibinfo
  {volume} {100}},\ \bibinfo {pages} {015010} (\bibinfo {year}
  {2019})}\BibitemShut {NoStop}%
\bibitem [{\citenamefont {Karshenboim}(2009)}]{Karshenboim:2008zj}%
  \BibitemOpen
  \bibfield  {author} {\bibinfo {author} {\bibfnamefont {S.~G.}\ \bibnamefont
  {Karshenboim}},\ }\href {\doibase 10.1134/S1063773709100028} {\bibfield
  {journal} {\bibinfo  {journal} {Astron. Lett.}\ }\textbf {\bibinfo {volume}
  {35}},\ \bibinfo {pages} {663} (\bibinfo {year} {2009})},\ \Eprint
  {http://arxiv.org/abs/0811.1008} {arXiv:0811.1008 [gr-qc]} \BibitemShut
  {NoStop}%
\bibitem [{\citenamefont {Stadnik}(2018)}]{Stadnik:2017yge}%
  \BibitemOpen
  \bibfield  {author} {\bibinfo {author} {\bibfnamefont {Y.~V.}\ \bibnamefont
  {Stadnik}},\ }\href {\doibase 10.1103/PhysRevLett.120.223202} {\bibfield
  {journal} {\bibinfo  {journal} {Phys. Rev. Lett.}\ }\textbf {\bibinfo
  {volume} {120}},\ \bibinfo {pages} {223202} (\bibinfo {year} {2018})},\
  \Eprint {http://arxiv.org/abs/1711.03700} {arXiv:1711.03700
  [physics.atom-ph]} \BibitemShut {NoStop}%
\bibitem [{\citenamefont {Gurung}\ \emph {et~al.}(2020)\citenamefont {Gurung},
  \citenamefont {Babij}, \citenamefont {Hogan},\ and\ \citenamefont
  {Cassidy}}]{2020-PS}%
  \BibitemOpen
  \bibfield  {author} {\bibinfo {author} {\bibfnamefont {L.}~\bibnamefont
  {Gurung}}, \bibinfo {author} {\bibfnamefont {T.~J.}\ \bibnamefont {Babij}},
  \bibinfo {author} {\bibfnamefont {S.~D.}\ \bibnamefont {Hogan}}, \ and\
  \bibinfo {author} {\bibfnamefont {D.~B.}\ \bibnamefont {Cassidy}},\ }\href
  {\doibase 10.1103/PhysRevLett.125.073002} {\bibfield  {journal} {\bibinfo
  {journal} {Phys. Rev. Lett.}\ }\textbf {\bibinfo {volume} {125}},\ \bibinfo
  {pages} {073002} (\bibinfo {year} {2020})}\BibitemShut {NoStop}%
\bibitem [{\citenamefont {Gurung}\ \emph {et~al.}(2021)\citenamefont {Gurung},
  \citenamefont {Babij}, \citenamefont {P\'erez-R\'{\i}os}, \citenamefont
  {Hogan},\ and\ \citenamefont {Cassidy}}]{2021-Ps}%
  \BibitemOpen
  \bibfield  {author} {\bibinfo {author} {\bibfnamefont {L.}~\bibnamefont
  {Gurung}}, \bibinfo {author} {\bibfnamefont {T.~J.}\ \bibnamefont {Babij}},
  \bibinfo {author} {\bibfnamefont {J.}~\bibnamefont {P\'erez-R\'{\i}os}},
  \bibinfo {author} {\bibfnamefont {S.~D.}\ \bibnamefont {Hogan}}, \ and\
  \bibinfo {author} {\bibfnamefont {D.~B.}\ \bibnamefont {Cassidy}},\ }\href
  {\doibase 10.1103/PhysRevA.103.042805} {\bibfield  {journal} {\bibinfo
  {journal} {Phys. Rev. A}\ }\textbf {\bibinfo {volume} {103}},\ \bibinfo
  {pages} {042805} (\bibinfo {year} {2021})}\BibitemShut {NoStop}%
\bibitem [{\citenamefont {Jungmann}(2016)}]{2016-Jung}%
  \BibitemOpen
  \bibfield  {author} {\bibinfo {author} {\bibfnamefont {K.~P.}\ \bibnamefont
  {Jungmann}},\ }\href {\doibase 10.7566/JPSJ.85.091004} {\bibfield  {journal}
  {\bibinfo  {journal} {Journal of the Physical Society of Japan}\ }\textbf
  {\bibinfo {volume} {85}},\ \bibinfo {pages} {091004} (\bibinfo {year}
  {2016})},\ \Eprint
  {http://arxiv.org/abs/https://doi.org/10.7566/JPSJ.85.091004}
  {https://doi.org/10.7566/JPSJ.85.091004} \BibitemShut {NoStop}%
\bibitem [{\citenamefont {Ohayon}\ \emph {et~al.}(2021)\citenamefont {Ohayon},
  \citenamefont {Burkley},\ and\ \citenamefont {Crivelli}}]{2021-MSPEC}%
  \BibitemOpen
  \bibfield  {author} {\bibinfo {author} {\bibfnamefont {B.}~\bibnamefont
  {Ohayon}}, \bibinfo {author} {\bibfnamefont {Z.}~\bibnamefont {Burkley}}, \
  and\ \bibinfo {author} {\bibfnamefont {P.}~\bibnamefont {Crivelli}},\ }\href
  {https://scipost.org/preprints/scipost_202106_00013v1} {\bibfield  {journal}
  {\bibinfo  {journal} {SciPost Physics Proceedings}\ } (\bibinfo {year}
  {2021})}\BibitemShut {NoStop}%
\bibitem [{\citenamefont {Bolton}\ \emph {et~al.}(1981)\citenamefont {Bolton},
  \citenamefont {Badertscher}, \citenamefont {Egan}, \citenamefont {Gardner},
  \citenamefont {Gladisch}, \citenamefont {Hughes}, \citenamefont {Lu},
  \citenamefont {Ritter}, \citenamefont {Souder}, \citenamefont {Vetter},
  \citenamefont {zu~Putlitz}, \citenamefont {Eckhause},\ and\ \citenamefont
  {Kane}}]{1981-M}%
  \BibitemOpen
  \bibfield  {author} {\bibinfo {author} {\bibfnamefont {P.~R.}\ \bibnamefont
  {Bolton}}, \bibinfo {author} {\bibfnamefont {A.}~\bibnamefont {Badertscher}},
  \bibinfo {author} {\bibfnamefont {P.~O.}\ \bibnamefont {Egan}}, \bibinfo
  {author} {\bibfnamefont {C.~J.}\ \bibnamefont {Gardner}}, \bibinfo {author}
  {\bibfnamefont {M.}~\bibnamefont {Gladisch}}, \bibinfo {author}
  {\bibfnamefont {V.~W.}\ \bibnamefont {Hughes}}, \bibinfo {author}
  {\bibfnamefont {D.~C.}\ \bibnamefont {Lu}}, \bibinfo {author} {\bibfnamefont
  {M.}~\bibnamefont {Ritter}}, \bibinfo {author} {\bibfnamefont {P.~A.}\
  \bibnamefont {Souder}}, \bibinfo {author} {\bibfnamefont {J.}~\bibnamefont
  {Vetter}}, \bibinfo {author} {\bibfnamefont {G.}~\bibnamefont {zu~Putlitz}},
  \bibinfo {author} {\bibfnamefont {M.}~\bibnamefont {Eckhause}}, \ and\
  \bibinfo {author} {\bibfnamefont {J.}~\bibnamefont {Kane}},\ }\href {\doibase
  10.1103/PhysRevLett.47.1441} {\bibfield  {journal} {\bibinfo  {journal}
  {Phys. Rev. Lett.}\ }\textbf {\bibinfo {volume} {47}},\ \bibinfo {pages}
  {1441} (\bibinfo {year} {1981})}\BibitemShut {NoStop}%
\bibitem [{\citenamefont {Oram}\ \emph {et~al.}(1984)\citenamefont {Oram},
  \citenamefont {Bailey}, \citenamefont {Schmor}, \citenamefont {Fry},
  \citenamefont {Kiefl}, \citenamefont {Warren}, \citenamefont {Marshall},\
  and\ \citenamefont {Olin}}]{1984-TRIUMF}%
  \BibitemOpen
  \bibfield  {author} {\bibinfo {author} {\bibfnamefont {C.~J.}\ \bibnamefont
  {Oram}}, \bibinfo {author} {\bibfnamefont {J.~M.}\ \bibnamefont {Bailey}},
  \bibinfo {author} {\bibfnamefont {P.~W.}\ \bibnamefont {Schmor}}, \bibinfo
  {author} {\bibfnamefont {C.~A.}\ \bibnamefont {Fry}}, \bibinfo {author}
  {\bibfnamefont {R.~F.}\ \bibnamefont {Kiefl}}, \bibinfo {author}
  {\bibfnamefont {J.~B.}\ \bibnamefont {Warren}}, \bibinfo {author}
  {\bibfnamefont {G.~M.}\ \bibnamefont {Marshall}}, \ and\ \bibinfo {author}
  {\bibfnamefont {A.}~\bibnamefont {Olin}},\ }\href {\doibase
  10.1103/PhysRevLett.52.910} {\bibfield  {journal} {\bibinfo  {journal} {Phys.
  Rev. Lett.}\ }\textbf {\bibinfo {volume} {52}},\ \bibinfo {pages} {910}
  (\bibinfo {year} {1984})}\BibitemShut {NoStop}%
\bibitem [{\citenamefont {Woodle}\ \emph {et~al.}(1990)\citenamefont {Woodle},
  \citenamefont {Badertscher}, \citenamefont {Hughes}, \citenamefont {Lu},
  \citenamefont {Ritter}, \citenamefont {Gladisch}, \citenamefont {Orth},
  \citenamefont {Zu~Putlitz}, \citenamefont {Eckhause}, \citenamefont {Kane}
  \emph {et~al.}}]{1990-LAMPF}%
  \BibitemOpen
  \bibfield  {author} {\bibinfo {author} {\bibfnamefont {K.}~\bibnamefont
  {Woodle}}, \bibinfo {author} {\bibfnamefont {A.}~\bibnamefont {Badertscher}},
  \bibinfo {author} {\bibfnamefont {V.}~\bibnamefont {Hughes}}, \bibinfo
  {author} {\bibfnamefont {D.}~\bibnamefont {Lu}}, \bibinfo {author}
  {\bibfnamefont {M.}~\bibnamefont {Ritter}}, \bibinfo {author} {\bibfnamefont
  {M.}~\bibnamefont {Gladisch}}, \bibinfo {author} {\bibfnamefont
  {H.}~\bibnamefont {Orth}}, \bibinfo {author} {\bibfnamefont {G.}~\bibnamefont
  {Zu~Putlitz}}, \bibinfo {author} {\bibfnamefont {M.}~\bibnamefont
  {Eckhause}}, \bibinfo {author} {\bibfnamefont {J.}~\bibnamefont {Kane}},
  \emph {et~al.},\ }\href {\doibase 10.1103/PhysRevA.41.93} {\bibfield
  {journal} {\bibinfo  {journal} {Physical Review A}\ }\textbf {\bibinfo
  {volume} {41}},\ \bibinfo {pages} {93} (\bibinfo {year} {1990})}\BibitemShut
  {NoStop}%
\bibitem [{\citenamefont {Morenzoni}\ \emph {et~al.}(2000)\citenamefont
  {Morenzoni}, \citenamefont {Glückler}, \citenamefont {Prokscha},
  \citenamefont {Weber}, \citenamefont {Forgan}, \citenamefont {Jackson},
  \citenamefont {Luetkens}, \citenamefont {Niedermayer}, \citenamefont
  {Pleines}, \citenamefont {Birke}, \citenamefont {Hofer}, \citenamefont
  {Litterst}, \citenamefont {Riseman},\ and\ \citenamefont
  {Schatz}}]{2000-MuSR}%
  \BibitemOpen
  \bibfield  {author} {\bibinfo {author} {\bibfnamefont {E.}~\bibnamefont
  {Morenzoni}}, \bibinfo {author} {\bibfnamefont {H.}~\bibnamefont
  {Glückler}}, \bibinfo {author} {\bibfnamefont {T.}~\bibnamefont {Prokscha}},
  \bibinfo {author} {\bibfnamefont {H.}~\bibnamefont {Weber}}, \bibinfo
  {author} {\bibfnamefont {E.}~\bibnamefont {Forgan}}, \bibinfo {author}
  {\bibfnamefont {T.}~\bibnamefont {Jackson}}, \bibinfo {author} {\bibfnamefont
  {H.}~\bibnamefont {Luetkens}}, \bibinfo {author} {\bibfnamefont
  {C.}~\bibnamefont {Niedermayer}}, \bibinfo {author} {\bibfnamefont
  {M.}~\bibnamefont {Pleines}}, \bibinfo {author} {\bibfnamefont
  {M.}~\bibnamefont {Birke}}, \bibinfo {author} {\bibfnamefont
  {A.}~\bibnamefont {Hofer}}, \bibinfo {author} {\bibfnamefont
  {J.}~\bibnamefont {Litterst}}, \bibinfo {author} {\bibfnamefont
  {T.}~\bibnamefont {Riseman}}, \ and\ \bibinfo {author} {\bibfnamefont
  {G.}~\bibnamefont {Schatz}},\ }\href {\doibase
  https://doi.org/10.1016/S0921-4526(00)00303-3} {\bibfield  {journal}
  {\bibinfo  {journal} {Physica B: Condensed Matter}\ }\textbf {\bibinfo
  {volume} {289-290}},\ \bibinfo {pages} {653} (\bibinfo {year}
  {2000})}\BibitemShut {NoStop}%
\bibitem [{\citenamefont {Prokscha}\ \emph {et~al.}(2008)\citenamefont
  {Prokscha}, \citenamefont {Morenzoni}, \citenamefont {Deiters}, \citenamefont
  {Foroughi}, \citenamefont {George}, \citenamefont {Kobler}, \citenamefont
  {Suter},\ and\ \citenamefont {Vrankovic}}]{2008-uE4}%
  \BibitemOpen
  \bibfield  {author} {\bibinfo {author} {\bibfnamefont {T.}~\bibnamefont
  {Prokscha}}, \bibinfo {author} {\bibfnamefont {E.}~\bibnamefont {Morenzoni}},
  \bibinfo {author} {\bibfnamefont {K.}~\bibnamefont {Deiters}}, \bibinfo
  {author} {\bibfnamefont {F.}~\bibnamefont {Foroughi}}, \bibinfo {author}
  {\bibfnamefont {D.}~\bibnamefont {George}}, \bibinfo {author} {\bibfnamefont
  {R.}~\bibnamefont {Kobler}}, \bibinfo {author} {\bibfnamefont
  {A.}~\bibnamefont {Suter}}, \ and\ \bibinfo {author} {\bibfnamefont
  {V.}~\bibnamefont {Vrankovic}},\ }\href {\doibase
  https://doi.org/10.1016/j.nima.2008.07.081} {\bibfield  {journal} {\bibinfo
  {journal} {Nuclear Instruments and Methods in Physics Research Section A:
  Accelerators, Spectrometers, Detectors and Associated Equipment}\ }\textbf
  {\bibinfo {volume} {595}},\ \bibinfo {pages} {317 } (\bibinfo {year}
  {2008})}\BibitemShut {NoStop}%
\bibitem [{\citenamefont {Janka}\ \emph {et~al.}(2020)\citenamefont {Janka},
  \citenamefont {Ohayon}, \citenamefont {Burkley}, \citenamefont {Gerchow},
  \citenamefont {Kuroda}, \citenamefont {Ni}, \citenamefont {Nishi},
  \citenamefont {Salman}, \citenamefont {Suter}, \citenamefont {Tuzi} \emph
  {et~al.}}]{2020-2S}%
  \BibitemOpen
  \bibfield  {author} {\bibinfo {author} {\bibfnamefont {G.}~\bibnamefont
  {Janka}}, \bibinfo {author} {\bibfnamefont {B.}~\bibnamefont {Ohayon}},
  \bibinfo {author} {\bibfnamefont {Z.}~\bibnamefont {Burkley}}, \bibinfo
  {author} {\bibfnamefont {L.}~\bibnamefont {Gerchow}}, \bibinfo {author}
  {\bibfnamefont {N.}~\bibnamefont {Kuroda}}, \bibinfo {author} {\bibfnamefont
  {X.}~\bibnamefont {Ni}}, \bibinfo {author} {\bibfnamefont {R.}~\bibnamefont
  {Nishi}}, \bibinfo {author} {\bibfnamefont {Z.}~\bibnamefont {Salman}},
  \bibinfo {author} {\bibfnamefont {A.}~\bibnamefont {Suter}}, \bibinfo
  {author} {\bibfnamefont {M.}~\bibnamefont {Tuzi}},  \emph {et~al.},\ }\href
  {\doibase https://doi.org/10.1140/epjc/s10052-020-8400-1} {\bibfield
  {journal} {\bibinfo  {journal} {The European Physical Journal C}\ }\textbf
  {\bibinfo {volume} {80}},\ \bibinfo {pages} {1} (\bibinfo {year}
  {2020})}\BibitemShut {NoStop}%
\bibitem [{\citenamefont {Salman}\ \emph {et~al.}(2012)\citenamefont {Salman},
  \citenamefont {Prokscha}, \citenamefont {Keller}, \citenamefont {Morenzoni},
  \citenamefont {Saadaoui}, \citenamefont {Sedlak}, \citenamefont {Shiroka},
  \citenamefont {Sidorov}, \citenamefont {Suter}, \citenamefont {Vrankovic},\
  and\ \citenamefont {Weber}}]{2012-SpinRot}%
  \BibitemOpen
  \bibfield  {author} {\bibinfo {author} {\bibfnamefont {Z.}~\bibnamefont
  {Salman}}, \bibinfo {author} {\bibfnamefont {T.}~\bibnamefont {Prokscha}},
  \bibinfo {author} {\bibfnamefont {P.}~\bibnamefont {Keller}}, \bibinfo
  {author} {\bibfnamefont {E.}~\bibnamefont {Morenzoni}}, \bibinfo {author}
  {\bibfnamefont {H.}~\bibnamefont {Saadaoui}}, \bibinfo {author}
  {\bibfnamefont {K.}~\bibnamefont {Sedlak}}, \bibinfo {author} {\bibfnamefont
  {T.}~\bibnamefont {Shiroka}}, \bibinfo {author} {\bibfnamefont
  {S.}~\bibnamefont {Sidorov}}, \bibinfo {author} {\bibfnamefont
  {A.}~\bibnamefont {Suter}}, \bibinfo {author} {\bibfnamefont
  {V.}~\bibnamefont {Vrankovic}}, \ and\ \bibinfo {author} {\bibfnamefont
  {H.-P.}\ \bibnamefont {Weber}},\ }\href {\doibase
  https://doi.org/10.1016/j.phpro.2012.04.039} {\bibfield  {journal} {\bibinfo
  {journal} {Physics Procedia}\ }\textbf {\bibinfo {volume} {30}},\ \bibinfo
  {pages} {55} (\bibinfo {year} {2012})},\ \bibinfo {note} {12th International
  Conference on Muon Spin Rotation, Relaxation and Resonance
  (muSR2011)}\BibitemShut {NoStop}%
\bibitem [{\citenamefont {Xiao}\ \emph {et~al.}(2017)\citenamefont {Xiao},
  \citenamefont {Morenzoni}, \citenamefont {Salman}, \citenamefont {Ye},\ and\
  \citenamefont {Prokscha}}]{xiao:2017nst}%
  \BibitemOpen
  \bibfield  {author} {\bibinfo {author} {\bibfnamefont {R.}~\bibnamefont
  {Xiao}}, \bibinfo {author} {\bibfnamefont {E.}~\bibnamefont {Morenzoni}},
  \bibinfo {author} {\bibfnamefont {Z.}~\bibnamefont {Salman}}, \bibinfo
  {author} {\bibfnamefont {B.-J.}\ \bibnamefont {Ye}}, \ and\ \bibinfo {author}
  {\bibfnamefont {T.}~\bibnamefont {Prokscha}},\ }\href {\doibase
  10.1007/s41365-017-0190-2} {\bibfield  {journal} {\bibinfo  {journal}
  {Nuclear Science and Techniques}\ }\textbf {\bibinfo {volume} {28}},\
  \bibinfo {pages} {29} (\bibinfo {year} {2017})}\BibitemShut {NoStop}%
\bibitem [{\citenamefont {Allegrini}\ \emph {et~al.}(2016)\citenamefont
  {Allegrini}, \citenamefont {Ebert},\ and\ \citenamefont
  {Funsten}}]{2016-Foils}%
  \BibitemOpen
  \bibfield  {author} {\bibinfo {author} {\bibfnamefont {F.}~\bibnamefont
  {Allegrini}}, \bibinfo {author} {\bibfnamefont {R.~W.}\ \bibnamefont
  {Ebert}}, \ and\ \bibinfo {author} {\bibfnamefont {H.~O.}\ \bibnamefont
  {Funsten}},\ }\href {\doibase https://doi.org/10.1002/2016JA022570}
  {\bibfield  {journal} {\bibinfo  {journal} {Journal of Geophysical Research:
  Space Physics}\ }\textbf {\bibinfo {volume} {121}},\ \bibinfo {pages} {3931}
  (\bibinfo {year} {2016})}\BibitemShut {NoStop}%
\bibitem [{\citenamefont {Yerokhin}\ \emph {et~al.}(2019)\citenamefont
  {Yerokhin}, \citenamefont {Pachucki},\ and\ \citenamefont
  {Patkóš}}]{2019-Pachuski}%
  \BibitemOpen
  \bibfield  {author} {\bibinfo {author} {\bibfnamefont {V.~A.}\ \bibnamefont
  {Yerokhin}}, \bibinfo {author} {\bibfnamefont {K.}~\bibnamefont {Pachucki}},
  \ and\ \bibinfo {author} {\bibfnamefont {V.}~\bibnamefont {Patkóš}},\
  }\href {\doibase 10.1002/andp.201800324} {\bibfield  {journal} {\bibinfo
  {journal} {Annalen der Physik}\ }\textbf {\bibinfo {volume} {531}},\ \bibinfo
  {pages} {1800324} (\bibinfo {year} {2019})}\BibitemShut {NoStop}%
\bibitem [{\citenamefont {Lundeen}\ and\ \citenamefont
  {Pipkin}(1986)}]{1986-LP}%
  \BibitemOpen
  \bibfield  {author} {\bibinfo {author} {\bibfnamefont {S.~R.}\ \bibnamefont
  {Lundeen}}\ and\ \bibinfo {author} {\bibfnamefont {F.~M.}\ \bibnamefont
  {Pipkin}},\ }\href {\doibase 10.1088/0026-1394/22/1/003} {\bibfield
  {journal} {\bibinfo  {journal} {Metrologia}\ }\textbf {\bibinfo {volume}
  {22}},\ \bibinfo {pages} {9} (\bibinfo {year} {1986})}\BibitemShut {NoStop}%
\bibitem [{\citenamefont {Sedlak}\ \emph {et~al.}(2012)\citenamefont {Sedlak},
  \citenamefont {Scheuermann}, \citenamefont {Shiroka}, \citenamefont
  {Stoykov}, \citenamefont {Raselli},\ and\ \citenamefont
  {Amato}}]{sedlak:musrsim2012}%
  \BibitemOpen
  \bibfield  {author} {\bibinfo {author} {\bibfnamefont {K.}~\bibnamefont
  {Sedlak}}, \bibinfo {author} {\bibfnamefont {R.}~\bibnamefont {Scheuermann}},
  \bibinfo {author} {\bibfnamefont {T.}~\bibnamefont {Shiroka}}, \bibinfo
  {author} {\bibfnamefont {A.}~\bibnamefont {Stoykov}}, \bibinfo {author}
  {\bibfnamefont {A.}~\bibnamefont {Raselli}}, \ and\ \bibinfo {author}
  {\bibfnamefont {A.}~\bibnamefont {Amato}},\ }\href {\doibase
  10.1016/j.phpro.2012.04.040} {\bibfield  {journal} {\bibinfo  {journal}
  {Physics Procedia}\ }\textbf {\bibinfo {volume} {30}},\ \bibinfo {pages} {61}
  (\bibinfo {year} {2012})}\BibitemShut {NoStop}%
\bibitem [{\citenamefont {Khaw}\ \emph {et~al.}(2015)\citenamefont {Khaw},
  \citenamefont {Antognini}, \citenamefont {Crivelli}, \citenamefont {Kirch},
  \citenamefont {Morenzoni}, \citenamefont {Salman}, \citenamefont {Suter},\
  and\ \citenamefont {Prokscha}}]{Khaw:2015eya}%
  \BibitemOpen
  \bibfield  {author} {\bibinfo {author} {\bibfnamefont {K.~S.}\ \bibnamefont
  {Khaw}}, \bibinfo {author} {\bibfnamefont {A.}~\bibnamefont {Antognini}},
  \bibinfo {author} {\bibfnamefont {P.}~\bibnamefont {Crivelli}}, \bibinfo
  {author} {\bibfnamefont {K.}~\bibnamefont {Kirch}}, \bibinfo {author}
  {\bibfnamefont {E.}~\bibnamefont {Morenzoni}}, \bibinfo {author}
  {\bibfnamefont {Z.}~\bibnamefont {Salman}}, \bibinfo {author} {\bibfnamefont
  {A.}~\bibnamefont {Suter}}, \ and\ \bibinfo {author} {\bibfnamefont
  {T.}~\bibnamefont {Prokscha}},\ }\href {\doibase
  10.1088/1748-0221/10/10/P10025} {\bibfield  {journal} {\bibinfo  {journal}
  {JINST}\ }\textbf {\bibinfo {volume} {10}},\ \bibinfo {pages} {P10025}
  (\bibinfo {year} {2015})}\BibitemShut {NoStop}%
\bibitem [{\citenamefont {Haas}\ \emph {et~al.}(2006)\citenamefont {Haas},
  \citenamefont {Jentschura}, \citenamefont {Keitel}, \citenamefont
  {Kolachevsky}, \citenamefont {Herrmann}, \citenamefont {Fendel},
  \citenamefont {Fischer}, \citenamefont {Udem}, \citenamefont {Holzwarth},
  \citenamefont {H\"ansch}, \citenamefont {Scully},\ and\ \citenamefont
  {Agarwal}}]{2006-Haas}%
  \BibitemOpen
  \bibfield  {author} {\bibinfo {author} {\bibfnamefont {M.}~\bibnamefont
  {Haas}}, \bibinfo {author} {\bibfnamefont {U.~D.}\ \bibnamefont
  {Jentschura}}, \bibinfo {author} {\bibfnamefont {C.~H.}\ \bibnamefont
  {Keitel}}, \bibinfo {author} {\bibfnamefont {N.}~\bibnamefont {Kolachevsky}},
  \bibinfo {author} {\bibfnamefont {M.}~\bibnamefont {Herrmann}}, \bibinfo
  {author} {\bibfnamefont {P.}~\bibnamefont {Fendel}}, \bibinfo {author}
  {\bibfnamefont {M.}~\bibnamefont {Fischer}}, \bibinfo {author} {\bibfnamefont
  {T.}~\bibnamefont {Udem}}, \bibinfo {author} {\bibfnamefont {R.}~\bibnamefont
  {Holzwarth}}, \bibinfo {author} {\bibfnamefont {T.~W.}\ \bibnamefont
  {H\"ansch}}, \bibinfo {author} {\bibfnamefont {M.~O.}\ \bibnamefont
  {Scully}}, \ and\ \bibinfo {author} {\bibfnamefont {G.~S.}\ \bibnamefont
  {Agarwal}},\ }\href {\doibase 10.1103/PhysRevA.73.052501} {\bibfield
  {journal} {\bibinfo  {journal} {Phys. Rev. A}\ }\textbf {\bibinfo {volume}
  {73}},\ \bibinfo {pages} {052501} (\bibinfo {year} {2006})}\BibitemShut
  {NoStop}%
\bibitem [{\citenamefont {Brun}\ and\ \citenamefont
  {Rademakers}(1997)}]{1997-ROOT}%
  \BibitemOpen
  \bibfield  {author} {\bibinfo {author} {\bibfnamefont {R.}~\bibnamefont
  {Brun}}\ and\ \bibinfo {author} {\bibfnamefont {F.}~\bibnamefont
  {Rademakers}},\ }\href {\doibase
  https://doi.org/10.1016/S0168-9002(97)00048-X} {\bibfield  {journal}
  {\bibinfo  {journal} {Nuclear Instruments and Methods in Physics Research
  Section A: Accelerators, Spectrometers, Detectors and Associated Equipment}\
  }\textbf {\bibinfo {volume} {389}},\ \bibinfo {pages} {81} (\bibinfo {year}
  {1997})},\ \bibinfo {note} {new Computing Techniques in Physics Research
  V}\BibitemShut {NoStop}%
\bibitem [{\citenamefont {Andrews}\ and\ \citenamefont
  {Newton}(1976{\natexlab{a}})}]{1976-Newton}%
  \BibitemOpen
  \bibfield  {author} {\bibinfo {author} {\bibfnamefont {D.~A.}\ \bibnamefont
  {Andrews}}\ and\ \bibinfo {author} {\bibfnamefont {G.}~\bibnamefont
  {Newton}},\ }\href {\doibase 10.1103/PhysRevLett.37.1254} {\bibfield
  {journal} {\bibinfo  {journal} {Phys. Rev. Lett.}\ }\textbf {\bibinfo
  {volume} {37}},\ \bibinfo {pages} {1254} (\bibinfo {year}
  {1976}{\natexlab{a}})}\BibitemShut {NoStop}%
\bibitem [{\citenamefont {Newton}\ \emph {et~al.}(1979)\citenamefont {Newton},
  \citenamefont {Andrews},\ and\ \citenamefont {Unsworth}}]{1979-Newton}%
  \BibitemOpen
  \bibfield  {author} {\bibinfo {author} {\bibfnamefont {G.}~\bibnamefont
  {Newton}}, \bibinfo {author} {\bibfnamefont {D.}~\bibnamefont {Andrews}}, \
  and\ \bibinfo {author} {\bibfnamefont {P.}~\bibnamefont {Unsworth}},\
  }\href@noop {} {\bibfield  {journal} {\bibinfo  {journal} {Philosophical
  Transactions of the Royal Society of London. Series A, Mathematical and
  Physical Sciences}\ }\textbf {\bibinfo {volume} {290}},\ \bibinfo {pages}
  {373} (\bibinfo {year} {1979})}\BibitemShut {NoStop}%
\bibitem [{\citenamefont {Andrews}\ and\ \citenamefont
  {Newton}(1976{\natexlab{b}})}]{1976-BS}%
  \BibitemOpen
  \bibfield  {author} {\bibinfo {author} {\bibfnamefont {D.~A.}\ \bibnamefont
  {Andrews}}\ and\ \bibinfo {author} {\bibfnamefont {G.}~\bibnamefont
  {Newton}},\ }\href {\doibase 10.1088/0022-3700/9/9/010} {\bibfield  {journal}
  {\bibinfo  {journal} {Journal of Physics B: Atomic and Molecular Physics}\
  }\textbf {\bibinfo {volume} {9}},\ \bibinfo {pages} {1453} (\bibinfo {year}
  {1976}{\natexlab{b}})}\BibitemShut {NoStop}%
\bibitem [{\citenamefont {Andrews}\ and\ \citenamefont
  {Newton}(1975)}]{1975-BS-EXP}%
  \BibitemOpen
  \bibfield  {author} {\bibinfo {author} {\bibfnamefont {D.~A.}\ \bibnamefont
  {Andrews}}\ and\ \bibinfo {author} {\bibfnamefont {G.}~\bibnamefont
  {Newton}},\ }\href {\doibase 10.1088/0022-3700/8/9/008} {\bibfield  {journal}
  {\bibinfo  {journal} {Journal of Physics B: Atomic and Molecular Physics}\
  }\textbf {\bibinfo {volume} {8}},\ \bibinfo {pages} {1415} (\bibinfo {year}
  {1975})}\BibitemShut {NoStop}%
\bibitem [{\citenamefont {Marsman}\ \emph {et~al.}(2017)\citenamefont
  {Marsman}, \citenamefont {Horbatsch},\ and\ \citenamefont
  {Hessels}}]{2017-QI}%
  \BibitemOpen
  \bibfield  {author} {\bibinfo {author} {\bibfnamefont {A.}~\bibnamefont
  {Marsman}}, \bibinfo {author} {\bibfnamefont {M.}~\bibnamefont {Horbatsch}},
  \ and\ \bibinfo {author} {\bibfnamefont {E.~A.}\ \bibnamefont {Hessels}},\
  }\href {\doibase 10.1103/PhysRevA.96.062111} {\bibfield  {journal} {\bibinfo
  {journal} {Phys. Rev. A}\ }\textbf {\bibinfo {volume} {96}},\ \bibinfo
  {pages} {062111} (\bibinfo {year} {2017})}\BibitemShut {NoStop}%
\bibitem [{\citenamefont {Peset}\ and\ \citenamefont
  {Pineda}(2015)}]{Peset:2015zga}%
  \BibitemOpen
  \bibfield  {author} {\bibinfo {author} {\bibfnamefont {C.}~\bibnamefont
  {Peset}}\ and\ \bibinfo {author} {\bibfnamefont {A.}~\bibnamefont {Pineda}},\
  }\href {\doibase 10.1140/epja/i2015-15156-2} {\bibfield  {journal} {\bibinfo
  {journal} {Eur. Phys. J. A}\ }\textbf {\bibinfo {volume} {51}},\ \bibinfo
  {pages} {156} (\bibinfo {year} {2015})},\ \Eprint
  {http://arxiv.org/abs/1508.01948} {arXiv:1508.01948 [hep-ph]} \BibitemShut
  {NoStop}%
\bibitem [{\citenamefont {Meyer}\ \emph {et~al.}(2000)\citenamefont {Meyer},
  \citenamefont {Bagayev}, \citenamefont {Baird}, \citenamefont {Bakule},
  \citenamefont {Boshier}, \citenamefont {Breitr\"uck}, \citenamefont
  {Cornish}, \citenamefont {Dychkov}, \citenamefont {Eaton}, \citenamefont
  {Grossmann}, \citenamefont {H\"ubl}, \citenamefont {Hughes}, \citenamefont
  {Jungmann}, \citenamefont {Lane}, \citenamefont {Liu}, \citenamefont {Lucas},
  \citenamefont {Matyugin}, \citenamefont {Merkel}, \citenamefont {zu~Putlitz},
  \citenamefont {Reinhard}, \citenamefont {Sandars}, \citenamefont {Santra},
  \citenamefont {Schmidt}, \citenamefont {Scott}, \citenamefont {Toner},
  \citenamefont {Towrie}, \citenamefont {Tr\"ager}, \citenamefont {Willmann},\
  and\ \citenamefont {Yakhontov}}]{2000-1S2S}%
  \BibitemOpen
  \bibfield  {author} {\bibinfo {author} {\bibfnamefont {V.}~\bibnamefont
  {Meyer}}, \bibinfo {author} {\bibfnamefont {S.~N.}\ \bibnamefont {Bagayev}},
  \bibinfo {author} {\bibfnamefont {P.~E.~G.}\ \bibnamefont {Baird}}, \bibinfo
  {author} {\bibfnamefont {P.}~\bibnamefont {Bakule}}, \bibinfo {author}
  {\bibfnamefont {M.~G.}\ \bibnamefont {Boshier}}, \bibinfo {author}
  {\bibfnamefont {A.}~\bibnamefont {Breitr\"uck}}, \bibinfo {author}
  {\bibfnamefont {S.~L.}\ \bibnamefont {Cornish}}, \bibinfo {author}
  {\bibfnamefont {S.}~\bibnamefont {Dychkov}}, \bibinfo {author} {\bibfnamefont
  {G.~H.}\ \bibnamefont {Eaton}}, \bibinfo {author} {\bibfnamefont
  {A.}~\bibnamefont {Grossmann}}, \bibinfo {author} {\bibfnamefont
  {D.}~\bibnamefont {H\"ubl}}, \bibinfo {author} {\bibfnamefont {V.~W.}\
  \bibnamefont {Hughes}}, \bibinfo {author} {\bibfnamefont {K.}~\bibnamefont
  {Jungmann}}, \bibinfo {author} {\bibfnamefont {I.~C.}\ \bibnamefont {Lane}},
  \bibinfo {author} {\bibfnamefont {Y.-W.}\ \bibnamefont {Liu}}, \bibinfo
  {author} {\bibfnamefont {D.}~\bibnamefont {Lucas}}, \bibinfo {author}
  {\bibfnamefont {Y.}~\bibnamefont {Matyugin}}, \bibinfo {author}
  {\bibfnamefont {J.}~\bibnamefont {Merkel}}, \bibinfo {author} {\bibfnamefont
  {G.}~\bibnamefont {zu~Putlitz}}, \bibinfo {author} {\bibfnamefont
  {I.}~\bibnamefont {Reinhard}}, \bibinfo {author} {\bibfnamefont {P.~G.~H.}\
  \bibnamefont {Sandars}}, \bibinfo {author} {\bibfnamefont {R.}~\bibnamefont
  {Santra}}, \bibinfo {author} {\bibfnamefont {P.~V.}\ \bibnamefont {Schmidt}},
  \bibinfo {author} {\bibfnamefont {C.~A.}\ \bibnamefont {Scott}}, \bibinfo
  {author} {\bibfnamefont {W.~T.}\ \bibnamefont {Toner}}, \bibinfo {author}
  {\bibfnamefont {M.}~\bibnamefont {Towrie}}, \bibinfo {author} {\bibfnamefont
  {K.}~\bibnamefont {Tr\"ager}}, \bibinfo {author} {\bibfnamefont
  {L.}~\bibnamefont {Willmann}}, \ and\ \bibinfo {author} {\bibfnamefont
  {V.}~\bibnamefont {Yakhontov}},\ }\href {\doibase
  10.1103/PhysRevLett.84.1136} {\bibfield  {journal} {\bibinfo  {journal}
  {Phys. Rev. Lett.}\ }\textbf {\bibinfo {volume} {84}},\ \bibinfo {pages}
  {1136} (\bibinfo {year} {2000})}\BibitemShut {NoStop}%
\bibitem [{\citenamefont {Jegerlehner}\ and\ \citenamefont
  {Nyffeler}(2009)}]{JEGERLEHNER20091}%
  \BibitemOpen
  \bibfield  {author} {\bibinfo {author} {\bibfnamefont {F.}~\bibnamefont
  {Jegerlehner}}\ and\ \bibinfo {author} {\bibfnamefont {A.}~\bibnamefont
  {Nyffeler}},\ }\href {\doibase https://doi.org/10.1016/j.physrep.2009.04.003}
  {\bibfield  {journal} {\bibinfo  {journal} {Physics Reports}\ }\textbf
  {\bibinfo {volume} {477}},\ \bibinfo {pages} {1} (\bibinfo {year}
  {2009})}\BibitemShut {NoStop}%
\bibitem [{\citenamefont {Delaunay}\ \emph {et~al.}(2021)\citenamefont
  {Delaunay}, \citenamefont {Ohayon},\ and\ \citenamefont {Soreq}}]{2021-g-2}%
  \BibitemOpen
  \bibfield  {author} {\bibinfo {author} {\bibfnamefont {C.}~\bibnamefont
  {Delaunay}}, \bibinfo {author} {\bibfnamefont {B.}~\bibnamefont {Ohayon}}, \
  and\ \bibinfo {author} {\bibfnamefont {Y.}~\bibnamefont {Soreq}},\
  }\href@noop {} {} (\bibinfo {year} {2021}),\ \Eprint
  {http://arxiv.org/abs/2106.11998} {arXiv:2106.11998 [hep-ph]} \BibitemShut
  {NoStop}%
\bibitem [{\citenamefont {Balkin}\ \emph {et~al.}(2021)\citenamefont {Balkin},
  \citenamefont {Delaunay}, \citenamefont {Geller}, \citenamefont {Kajomovitz},
  \citenamefont {Perez}, \citenamefont {Shpilman},\ and\ \citenamefont
  {Soreq}}]{Balkin:2021rvh}%
  \BibitemOpen
  \bibfield  {author} {\bibinfo {author} {\bibfnamefont {R.}~\bibnamefont
  {Balkin}}, \bibinfo {author} {\bibfnamefont {C.}~\bibnamefont {Delaunay}},
  \bibinfo {author} {\bibfnamefont {M.}~\bibnamefont {Geller}}, \bibinfo
  {author} {\bibfnamefont {E.}~\bibnamefont {Kajomovitz}}, \bibinfo {author}
  {\bibfnamefont {G.}~\bibnamefont {Perez}}, \bibinfo {author} {\bibfnamefont
  {Y.}~\bibnamefont {Shpilman}}, \ and\ \bibinfo {author} {\bibfnamefont
  {Y.}~\bibnamefont {Soreq}},\ }\href@noop {} {\  (\bibinfo {year} {2021})},\
  \Eprint {http://arxiv.org/abs/2104.08289} {arXiv:2104.08289 [hep-ph]}
  \BibitemShut {NoStop}%
\bibitem [{\citenamefont {Fadeev}\ \emph {et~al.}(2019)\citenamefont {Fadeev},
  \citenamefont {Stadnik}, \citenamefont {Ficek}, \citenamefont {Kozlov},
  \citenamefont {Flambaum},\ and\ \citenamefont {Budker}}]{Fadeev:2018rfl}%
  \BibitemOpen
  \bibfield  {author} {\bibinfo {author} {\bibfnamefont {P.}~\bibnamefont
  {Fadeev}}, \bibinfo {author} {\bibfnamefont {Y.~V.}\ \bibnamefont {Stadnik}},
  \bibinfo {author} {\bibfnamefont {F.}~\bibnamefont {Ficek}}, \bibinfo
  {author} {\bibfnamefont {M.~G.}\ \bibnamefont {Kozlov}}, \bibinfo {author}
  {\bibfnamefont {V.~V.}\ \bibnamefont {Flambaum}}, \ and\ \bibinfo {author}
  {\bibfnamefont {D.}~\bibnamefont {Budker}},\ }\href {\doibase
  10.1103/PhysRevA.99.022113} {\bibfield  {journal} {\bibinfo  {journal} {Phys.
  Rev. A}\ }\textbf {\bibinfo {volume} {99}},\ \bibinfo {pages} {022113}
  (\bibinfo {year} {2019})},\ \Eprint {http://arxiv.org/abs/1810.10364}
  {arXiv:1810.10364 [hep-ph]} \BibitemShut {NoStop}%
\bibitem [{\citenamefont {Abi}\ \emph {et~al.}(2021)\citenamefont {Abi} \emph
  {et~al.}}]{Muong-2:2021ojo}%
  \BibitemOpen
  \bibfield  {author} {\bibinfo {author} {\bibfnamefont {B.}~\bibnamefont
  {Abi}} \emph {et~al.} (\bibinfo {collaboration} {Muon g-2}),\ }\href
  {\doibase 10.1103/PhysRevLett.126.141801} {\bibfield  {journal} {\bibinfo
  {journal} {Phys. Rev. Lett.}\ }\textbf {\bibinfo {volume} {126}},\ \bibinfo
  {pages} {141801} (\bibinfo {year} {2021})},\ \Eprint
  {http://arxiv.org/abs/2104.03281} {arXiv:2104.03281 [hep-ex]} \BibitemShut
  {NoStop}%
\bibitem [{\citenamefont {Hanneke}\ \emph {et~al.}(2008)\citenamefont
  {Hanneke}, \citenamefont {Fogwell},\ and\ \citenamefont
  {Gabrielse}}]{Hanneke:2008tm}%
  \BibitemOpen
  \bibfield  {author} {\bibinfo {author} {\bibfnamefont {D.}~\bibnamefont
  {Hanneke}}, \bibinfo {author} {\bibfnamefont {S.}~\bibnamefont {Fogwell}}, \
  and\ \bibinfo {author} {\bibfnamefont {G.}~\bibnamefont {Gabrielse}},\ }\href
  {\doibase 10.1103/PhysRevLett.100.120801} {\bibfield  {journal} {\bibinfo
  {journal} {Phys. Rev. Lett.}\ }\textbf {\bibinfo {volume} {100}},\ \bibinfo
  {pages} {120801} (\bibinfo {year} {2008})},\ \Eprint
  {http://arxiv.org/abs/0801.1134} {arXiv:0801.1134 [physics.atom-ph]}
  \BibitemShut {NoStop}%
\bibitem [{\citenamefont {Parker}\ \emph {et~al.}(2018)\citenamefont {Parker},
  \citenamefont {Yu}, \citenamefont {Zhong}, \citenamefont {Estey},\ and\
  \citenamefont {M\"uller}}]{Parker:2018vye}%
  \BibitemOpen
  \bibfield  {author} {\bibinfo {author} {\bibfnamefont {R.~H.}\ \bibnamefont
  {Parker}}, \bibinfo {author} {\bibfnamefont {C.}~\bibnamefont {Yu}}, \bibinfo
  {author} {\bibfnamefont {W.}~\bibnamefont {Zhong}}, \bibinfo {author}
  {\bibfnamefont {B.}~\bibnamefont {Estey}}, \ and\ \bibinfo {author}
  {\bibfnamefont {H.}~\bibnamefont {M\"uller}},\ }\href {\doibase
  10.1126/science.aap7706} {\bibfield  {journal} {\bibinfo  {journal}
  {Science}\ }\textbf {\bibinfo {volume} {360}},\ \bibinfo {pages} {191}
  (\bibinfo {year} {2018})},\ \Eprint {http://arxiv.org/abs/1812.04130}
  {arXiv:1812.04130 [physics.atom-ph]} \BibitemShut {NoStop}%
\bibitem [{\citenamefont {Morel}\ \emph {et~al.}(2020)\citenamefont {Morel},
  \citenamefont {Yao}, \citenamefont {Clad\'e},\ and\ \citenamefont
  {Guellati-Kh\'elifa}}]{Morel:2020dww}%
  \BibitemOpen
  \bibfield  {author} {\bibinfo {author} {\bibfnamefont {L.}~\bibnamefont
  {Morel}}, \bibinfo {author} {\bibfnamefont {Z.}~\bibnamefont {Yao}}, \bibinfo
  {author} {\bibfnamefont {P.}~\bibnamefont {Clad\'e}}, \ and\ \bibinfo
  {author} {\bibfnamefont {S.}~\bibnamefont {Guellati-Kh\'elifa}},\ }\href
  {\doibase 10.1038/s41586-020-2964-7} {\bibfield  {journal} {\bibinfo
  {journal} {Nature}\ }\textbf {\bibinfo {volume} {588}},\ \bibinfo {pages}
  {61} (\bibinfo {year} {2020})}\BibitemShut {NoStop}%
\bibitem [{\citenamefont {Andreev}\ \emph {et~al.}(2021)\citenamefont {Andreev}
  \emph {et~al.}}]{NA64:2021xzo}%
  \BibitemOpen
  \bibfield  {author} {\bibinfo {author} {\bibfnamefont {Y.~M.}\ \bibnamefont
  {Andreev}} \emph {et~al.} (\bibinfo {collaboration} {NA64}),\ }\href
  {\doibase 10.1103/PhysRevLett.126.211802} {\bibfield  {journal} {\bibinfo
  {journal} {Phys. Rev. Lett.}\ }\textbf {\bibinfo {volume} {126}},\ \bibinfo
  {pages} {211802} (\bibinfo {year} {2021})},\ \Eprint
  {http://arxiv.org/abs/2102.01885} {arXiv:2102.01885 [hep-ex]} \BibitemShut
  {NoStop}%
\bibitem [{\citenamefont {Allegrini}\ \emph {et~al.}(2014)\citenamefont
  {Allegrini}, \citenamefont {Ebert}, \citenamefont {Fuselier}, \citenamefont
  {Nicolaou}, \citenamefont {Bedworth}, \citenamefont {Sinton},\ and\
  \citenamefont {Trattner}}]{2014-GraphState}%
  \BibitemOpen
  \bibfield  {author} {\bibinfo {author} {\bibfnamefont {F.}~\bibnamefont
  {Allegrini}}, \bibinfo {author} {\bibfnamefont {R.~W.}\ \bibnamefont
  {Ebert}}, \bibinfo {author} {\bibfnamefont {S.~A.}\ \bibnamefont {Fuselier}},
  \bibinfo {author} {\bibfnamefont {G.}~\bibnamefont {Nicolaou}}, \bibinfo
  {author} {\bibfnamefont {P.~V.}\ \bibnamefont {Bedworth}}, \bibinfo {author}
  {\bibfnamefont {S.~W.}\ \bibnamefont {Sinton}}, \ and\ \bibinfo {author}
  {\bibfnamefont {K.~J.}\ \bibnamefont {Trattner}},\ }\href {\doibase
  10.1117/1.OE.53.2.024101} {\bibfield  {journal} {\bibinfo  {journal} {Optical
  Engineering}\ }\textbf {\bibinfo {volume} {53}},\ \bibinfo {pages} {1 }
  (\bibinfo {year} {2014})}\BibitemShut {NoStop}%
\bibitem [{\citenamefont {Vira}\ \emph {et~al.}(2020)\citenamefont {Vira},
  \citenamefont {Fernandes}, \citenamefont {Funsten}, \citenamefont {Morley},
  \citenamefont {Yamaguchi}, \citenamefont {Liu},\ and\ \citenamefont
  {Moody}}]{2020-Graph}%
  \BibitemOpen
  \bibfield  {author} {\bibinfo {author} {\bibfnamefont {A.~D.}\ \bibnamefont
  {Vira}}, \bibinfo {author} {\bibfnamefont {P.~A.}\ \bibnamefont {Fernandes}},
  \bibinfo {author} {\bibfnamefont {H.~O.}\ \bibnamefont {Funsten}}, \bibinfo
  {author} {\bibfnamefont {S.~K.}\ \bibnamefont {Morley}}, \bibinfo {author}
  {\bibfnamefont {H.}~\bibnamefont {Yamaguchi}}, \bibinfo {author}
  {\bibfnamefont {F.}~\bibnamefont {Liu}}, \ and\ \bibinfo {author}
  {\bibfnamefont {N.~A.}\ \bibnamefont {Moody}},\ }\href {\doibase
  10.1063/1.5134768} {\bibfield  {journal} {\bibinfo  {journal} {Review of
  Scientific Instruments}\ }\textbf {\bibinfo {volume} {91}},\ \bibinfo {pages}
  {033302} (\bibinfo {year} {2020})},\ \Eprint
  {http://arxiv.org/abs/https://doi.org/10.1063/1.5134768}
  {https://doi.org/10.1063/1.5134768} \BibitemShut {NoStop}%
\bibitem [{\citenamefont {Bezginov}\ \emph {et~al.}(2019)\citenamefont
  {Bezginov}, \citenamefont {Valdez}, \citenamefont {Horbatsch}, \citenamefont
  {Marsman}, \citenamefont {Vutha},\ and\ \citenamefont
  {Hessels}}]{2019-Hessels}%
  \BibitemOpen
  \bibfield  {author} {\bibinfo {author} {\bibfnamefont {N.}~\bibnamefont
  {Bezginov}}, \bibinfo {author} {\bibfnamefont {T.}~\bibnamefont {Valdez}},
  \bibinfo {author} {\bibfnamefont {M.}~\bibnamefont {Horbatsch}}, \bibinfo
  {author} {\bibfnamefont {A.}~\bibnamefont {Marsman}}, \bibinfo {author}
  {\bibfnamefont {A.~C.}\ \bibnamefont {Vutha}}, \ and\ \bibinfo {author}
  {\bibfnamefont {E.~A.}\ \bibnamefont {Hessels}},\ }\href {\doibase
  10.1126/science.aau7807} {\bibfield  {journal} {\bibinfo  {journal}
  {Science}\ }\textbf {\bibinfo {volume} {365}},\ \bibinfo {pages} {1007}
  (\bibinfo {year} {2019})}\BibitemShut {NoStop}%
\bibitem [{\citenamefont {Kirch}()}]{2017-HiMBMuCOOL}%
  \BibitemOpen
  \bibfield  {author} {\bibinfo {author} {\bibfnamefont {K.}~\bibnamefont
  {Kirch}},\ }\enquote {\bibinfo {title} {Slow muons and muonium},}\ in\ \href
  {\doibase 10.1142/9789813148505_0014} {\emph {\bibinfo {booktitle} {CPT and
  Lorentz Symmetry}}},\ pp.\ \bibinfo {pages} {53--56}\BibitemShut {NoStop}%
\bibitem [{\citenamefont {Crivelli}(2018)}]{2018-MuMASS}%
  \BibitemOpen
  \bibfield  {author} {\bibinfo {author} {\bibfnamefont {P.}~\bibnamefont
  {Crivelli}},\ }\href {\doibase 10.1007/s10751-018-1525-z} {\bibfield
  {journal} {\bibinfo  {journal} {Hyperfine Interactions}\ }\textbf {\bibinfo
  {volume} {239}},\ \bibinfo {pages} {49} (\bibinfo {year} {2018})}\BibitemShut
  {NoStop}%
\bibitem [{\citenamefont {Kanda}\ \emph {et~al.}(2021)\citenamefont {Kanda},
  \citenamefont {Fukao}, \citenamefont {Ikedo}, \citenamefont {Ishida},
  \citenamefont {Iwasaki}, \citenamefont {Kawall}, \citenamefont {Kawamura},
  \citenamefont {Kojima}, \citenamefont {Kurosawa}, \citenamefont {Matsuda},
  \citenamefont {Mibe}, \citenamefont {Miyake}, \citenamefont {Nishimura},
  \citenamefont {Saito}, \citenamefont {Sato}, \citenamefont {Seo},
  \citenamefont {Shimomura}, \citenamefont {Strasser}, \citenamefont {Tanaka},
  \citenamefont {Tanaka}, \citenamefont {Torii}, \citenamefont {Toyoda},\ and\
  \citenamefont {Ueno}}]{2020-MUSEUM}%
  \BibitemOpen
  \bibfield  {author} {\bibinfo {author} {\bibfnamefont {S.}~\bibnamefont
  {Kanda}}, \bibinfo {author} {\bibfnamefont {Y.}~\bibnamefont {Fukao}},
  \bibinfo {author} {\bibfnamefont {Y.}~\bibnamefont {Ikedo}}, \bibinfo
  {author} {\bibfnamefont {K.}~\bibnamefont {Ishida}}, \bibinfo {author}
  {\bibfnamefont {M.}~\bibnamefont {Iwasaki}}, \bibinfo {author} {\bibfnamefont
  {D.}~\bibnamefont {Kawall}}, \bibinfo {author} {\bibfnamefont
  {N.}~\bibnamefont {Kawamura}}, \bibinfo {author} {\bibfnamefont
  {K.}~\bibnamefont {Kojima}}, \bibinfo {author} {\bibfnamefont
  {N.}~\bibnamefont {Kurosawa}}, \bibinfo {author} {\bibfnamefont
  {Y.}~\bibnamefont {Matsuda}}, \bibinfo {author} {\bibfnamefont
  {T.}~\bibnamefont {Mibe}}, \bibinfo {author} {\bibfnamefont {Y.}~\bibnamefont
  {Miyake}}, \bibinfo {author} {\bibfnamefont {S.}~\bibnamefont {Nishimura}},
  \bibinfo {author} {\bibfnamefont {N.}~\bibnamefont {Saito}}, \bibinfo
  {author} {\bibfnamefont {Y.}~\bibnamefont {Sato}}, \bibinfo {author}
  {\bibfnamefont {S.}~\bibnamefont {Seo}}, \bibinfo {author} {\bibfnamefont
  {K.}~\bibnamefont {Shimomura}}, \bibinfo {author} {\bibfnamefont
  {P.}~\bibnamefont {Strasser}}, \bibinfo {author} {\bibfnamefont
  {K.}~\bibnamefont {Tanaka}}, \bibinfo {author} {\bibfnamefont
  {T.}~\bibnamefont {Tanaka}}, \bibinfo {author} {\bibfnamefont
  {H.}~\bibnamefont {Torii}}, \bibinfo {author} {\bibfnamefont
  {A.}~\bibnamefont {Toyoda}}, \ and\ \bibinfo {author} {\bibfnamefont
  {Y.}~\bibnamefont {Ueno}},\ }\href {\doibase
  https://doi.org/10.1016/j.physletb.2021.136154} {\bibfield  {journal}
  {\bibinfo  {journal} {Physics Letters B}\ }\textbf {\bibinfo {volume}
  {815}},\ \bibinfo {pages} {136154} (\bibinfo {year} {2021})}\BibitemShut
  {NoStop}%
\bibitem [{\citenamefont {Hardy}\ and\ \citenamefont
  {Lasenby}(2017)}]{2017-Stellar}%
  \BibitemOpen
  \bibfield  {author} {\bibinfo {author} {\bibfnamefont {E.}~\bibnamefont
  {Hardy}}\ and\ \bibinfo {author} {\bibfnamefont {R.}~\bibnamefont
  {Lasenby}},\ }\href {\doibase https://doi.org/10.1007/JHEP02(2017)033}
  {\bibfield  {journal} {\bibinfo  {journal} {Journal of High Energy Physics}\
  }\textbf {\bibinfo {volume} {2017}},\ \bibinfo {pages} {1} (\bibinfo {year}
  {2017})}\BibitemShut {NoStop}%
\end{thebibliography}%


%

\end{document}